\documentclass[%
 reprint,
 showpacs,preprintnumbers,
 amsmath,amssymb,
 aps,
 prb,
]{revtex4-1}
\usepackage{dcolumn}
\usepackage{graphicx,color}
\usepackage{subfigure}
\usepackage{here}
\usepackage{mathrsfs}

\makeatletter
\def\btt#1{\texttt{\@backslashchar#1}}
\DeclareRobustCommand\bblash{\btt{\@backslashchar}} \makeatother

\begin{document}

\title{Three-orbital continuous model for $1H$-type metallic transition-metal dichalcogenide monolayers}

\author{Tetsuro Habe}
\affiliation{Nagamori Institute, Kyoto University of Advanced Science, Kyoto 615-0096, Japan\\
Department of Applied Physics, Hokkaido University, Sapporo, Hokkaido 060-0808, Japan}

\date{\today}

\begin{abstract}
We theoretically investigate the electronic states in monolayer NbSe$_2$ and develop continuous models to describe these states in Fermi pockets.
In $1H$-type metallic transition-metal dichalcogenides(TMDCs), the Femi surface consists of three pockets enclosing the $\Gamma$, $K$, and $K'$ points.
We reveal that the conventional effective model used for semiconducting TMDCs is not sufficient to describe the electronic states in metallic TMDCs and thus introduce a scheme to construct the effective model from the first-principles results.
All models can be represented by $3\times3$ Hamiltonian and well reproduce electronic states around the Fermi energy in terms of the orbital composition and the phase factor.
We also show that the $p$ orbitals in chalcogen atoms, which are ignored in the conventional $2\times2$ model, play a crucial role in metallic TMDCs.
Although the aim of these models is to reproduce electronic states, they can well describe states near the high-symmetry points and the profile of  Berry curvature in the wave vector space.
The continuous model can be a handleable tool to describe the electronic states and to analyze the transport phenomena in metallic TMDCs.
\end{abstract}

\maketitle
\section{Introduction}
Transition-metal dichalcogenide (TMDC) is a group of materials composed of transition-metal and chalcogen atoms.
The group includes many atomic layered materials consisting of different elements and provides several electrically different materials; metal\cite{Naito1982}, semiconductor\cite{Splendiani2010}, superconductor\cite{Lu2015,Xi2015-super}, and topological material\cite{Tang2017}.
The semiconducting $1H$-type monolayer TMDCs have been analyzed by using an effective model so-called massive Dirac Hamiltonian with the Zeeman-like spin-orbit coupling (SOC)\cite{Xiao2012} or that with some additional terms\cite{Kormanyos2015}.
The effective model has revealed fascinating phenomena in optics and electronics; valley-selective optical absorption\cite{Cao2012,Mak2012,Shi2013}, anomalous Hall effect\cite{habe2017}, quantized valley Hall effect\cite{Xiao2012,Kormanyos2018}, optically-induced spin Hall effect\cite{Shan2013}, and the electronic transport properties\cite{Ochoa2013-1,Ochoa2013-2,Song2013,Hatami2014,Kormanyos2014,Habe2015,Habe2016}.

Recently, metallic momolayer-TMDCs attract much attention in terms of electronic property.
This is because the metallic TMDC drastically changes the electronic property by reducing the number of layers.
For example, monolayer NbSe$_2$, a metallic TMDC, shows Ising supercondivity\cite{Lu2015,Xi2015-super,Xi2017,Wang2017,Sohn2018,Hamill2020} and the change of order in the charge density wave phase\cite{Ugeda2015,Xi2015,Zheng2018}.
These attractive phenomena have been analyzed by using the first-principles calculation and the effective model for semiconducting TMDCs, the modified Dirac model.\cite{David2018,Rahimi2018,Habe2019-1,Habe2019-2,Shaffer2019,Glodzik2019,Sticlet2019,Divilov2020,Hamill2020,Liu2020}
However, the first-principles calculation is hard to provide a simple picture behind the phenomena and the modified Dirac model does not well describe the electronic states in metallic TMDCs as discussed in this paper.

\begin{figure}[htbp]
\begin{center}
\includegraphics[width=85mm]{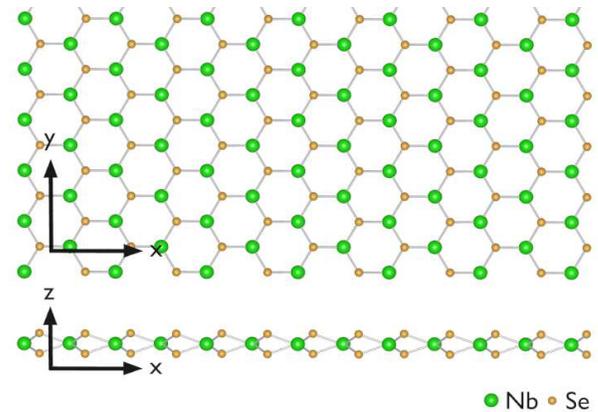}
\caption{
Schematics of monolayer crystal structure of NbSe$_2$. The upper and lower structures show the vertical view and the horizontal view, respectively. Other $1H$-type metallic TMDCs have the same structure with replacing Nb and Se by the transition-metal and chalcogen atoms, respectively.
 }\label{fig_schematic}
\end{center}
\end{figure}

In this paper, we introduce easily-handleable effective models, Eqs.\ (\ref{eq_hamiltonain_gamma}) and (\ref{eq_Hamiltonian_k}), for three Fermi pockets in $1H$-type metallic TMDCs.
These effective models are continuous in the wave vector space and adjusted to the first-principles band structure and the electronic states around the Fermi surface in the three valleys; the $\Gamma$, $K$, and $K'$ valleys in the first Brillouin zone.
We show that these models should be represented in the three-orbital basis, i.e., a $3\times3$ Hamiltonian, even though that for semiconducting TMDCs is defined in a two-orbital basis.
The three orbitals are composite orbitals of pure electronic orbitals in the transition-metal and chalcogen atoms.
We reveal that the $p$-orbitals play a crucial role to reproduce the electronic states although they have been ignored in conventional effective models.
We also present a scheme to obtain the parameter-set to reproduce the energy dispersion and the electronic states in monolayer NbSe$_2$ in Fig.\ \ref{fig_schematic} as an example.

\begin{figure}[htbp]
\begin{center}
\includegraphics[width=75mm]{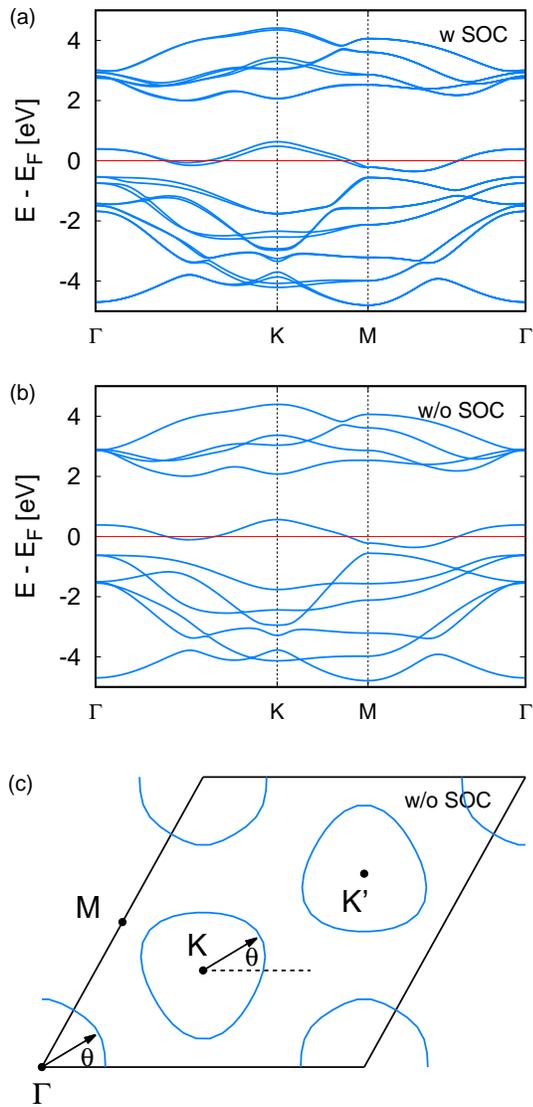}
\caption{
The electronic band structure of monolayer NbSe$_2$ with and without SOC in (a) and (b), respectively. The horizontal line indicates the Fermi energy.  The Fermi surface without the SOC effect is presented in (c).
 }\label{fig_band}
\end{center}
\end{figure} 
\section{First-principles band structure}\label{sec_first_principles}
Firstly, we show the first-principles electronic band structure of monolayer NbSe$_2$.
Here, we start with this specific material but the $1H$-type monolayer TMDCs have the similar band structure and electronic state to those of NbSe$_2$.
The band structure is obtained by using QUANTUM-ESPRESSO\cite{quantum-espresso}, a calculation code of density functional theory, with applying projector-argument-wave method and generalized gradient approximation functional.
Here, the energy cut-off is 50 Ry for the plane wave basis and 500 Ry for the charge density, and the convergence criterion is $10^{-8}$ Ry.
The lattice parameters are also computed by using the same code and estimated to be $c=3.475$\AA\ and $d=3.514$\AA\ as the lattice constant and the distance between the top and bottom sublayer of chalcogen atoms, respectively.
In Fig.\ \ref{fig_band}, the band structures (a) and (b) are obtained in the presence and absence of SOC, respectively.
In the Fermi surface, there are three disconnected pockets enclosing the $\Gamma$, $K$, and $K'$ points in the first Brillouin zone (see Fig.\ \ref{fig_band}(c)).
Since the electronic transport properties including the superconductivity are dominated by electronic states around the Fermi level, we develop effective models reproducing these states in terms of the energy dispersion and the atomic-orbital composition.
We start with constructing the effective model without SOC and then introduce SOC in the following section.

\begin{figure}[htbp]
\begin{center}
 \includegraphics[width=90mm]{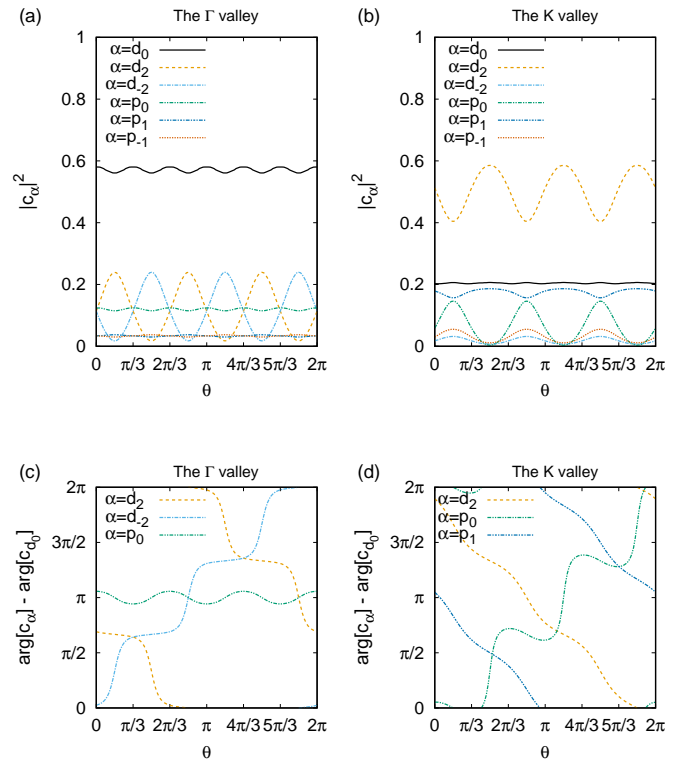}
\caption{
The orbital composition of electronic states in the Fermi pockets around the $\Gamma$, $K$, and $K'$ points. Here $d_m$ and $p_m$ indicate the $d$ and $p$ orbitals with the magnetic quantum number $m$, respectively.
 }\label{fig_orbitals}
\end{center}
\end{figure} 
\section{The effective model without SOC}
To construct the effective model, we calculate the atomic-orbital composition of electronic states in the Fermi surface from the first-principles band structure by using Wannier90\cite{wannier90}, which enables to compute the maximally-localized Wannier orbitals and the hopping matrix from the first-principles bands.
We adopt five $d$ orbitals in Nb atom and six $p$ orbitals in top and bottom Se atoms as the Wannier orbitals.
In the Fig.\ \ref{fig_orbitals}(a) and (b), the atomic composition of electronic states in the Fermi surface is presented.
Here each orbital is denoted by $d_\alpha$ or $p_\alpha$, where $d$ and $p$ are the labels of orbital and the subscript $\alpha$ implies the magnetic quantum number with respect to the $z$ axis.
The left and right panels are corresponding to the Fermi pockets around the $\Gamma$ and $K$ points, called the $\Gamma$ and $K$ valleys, respectively.
Each electronic state is labeled by the angle $\theta$ of wave vector with respect to the center, i.e., the high symmetry point, of the pocket as shown in Fig.\ \ref{fig_band}(c).
Here the results of the $K'$ valley can be obtained from those of the $K$ valley by the sign change as $\theta\rightarrow-\theta$ and $d_{\alpha}\rightarrow d_{-\alpha}$ because of time-reversal symmetry.
The conventional set of orbitals\cite{Xiao2012}, which adopted to construct the Dirac model in the case of semiconducting TMDC, is $d_0$ for the $\Gamma$ valley, ($d_0$, $d_2$) for the $K$ valley, and ($d_0$, $d_{-2}$) for the $K'$ valley. 
However the conventional sets account for 60\%-70\% of the amplitude in these realistic electronic states.
Thus the set of orbitals has to be rearranged for describing the electronic states in each valley and constructing effective models for the three valleys.

\subsection{The $\Gamma$ valley}
In the $\Gamma$ valley, we adopt four atomic orbitals; $p_0$,  $d_0$, and $d_{\pm2}$, to improve the reproducibility of orbital composition in electronic states.
The electronic states are represented by the superposition of these Wannier functions $|\psi_\alpha\rangle$,
\begin{align}
|\Psi\rangle=c_{d_0}|\psi_{d_0}\rangle+c_{p_0}|\psi_{p_0}\rangle+c_{d_2}|\psi_{d_2}\rangle+c_{d_{-2}}|\psi_{d_{-2}}\rangle,
\end{align}
where the variable $\boldsymbol{k}$ is omitted for the simple notation.
Here $|c_\alpha|^2$ is the amplitude corresponding to the quantity shown in Fig.\ \ref{fig_orbitals}(a).
For reproducing the electronic state $|\Psi\rangle$, the phase factor of coefficient $c_\alpha$ is also calculated and shown in Fig.\ \ref{fig_orbitals}(c).
The relative phase between the $d_2$ ($d_{-2}$) and $d_0$ orbitals decreases (increases) with the angle $\theta$ of wave vector.
Moreover, the amplitude ratio between the $d_0$ and $d_{\pm2}$ orbitals also varies with $\theta$.
The $p_0$ orbital, on the other hand, follows the $d_0$ orbital in terms of the phase factor and the amplitude.
Thus, we can introduce a composite orbital as a part of basis by mixing the $d_0$ orbital and the $p_0$ orbital defined as 
\begin{align}
|d_0+p_0\rangle=r_1|d_0\rangle+r_2|p_0\rangle,
\end{align}
where the coefficients are given by $r_1=\sqrt{0.77}$ and $r_2=-\sqrt{0.23}$ in the case of NbSe$_2$.
The three orbitals account for over 90\% of the amplitude of wave functions in the $\Gamma$ valley.
Therefore we adopt the three orbitals as the basis for our effective model and construct $3\times3$ Hamiltonian to describe the electronic states in this valley.

For the complete set, we introduce two fake bands to construct the effective model in the $\Gamma$ valley.
Although the partially-filled band, which is crossing the Fermi level, consists mostly of the three orbitals, other bands are not consisting of only the three orbitals but also the $p_{\pm1}$ orbitals with a non-negligible amplitude in the first-principles bands.
Thus it is impossible to reproduce these bands in terms of the orbital composition by using the three-orbital model.
Therefore we introduce two high energy bands, the fake bands, consisting only of the three orbitals instead of the realistic bands..
Since the state in the partially-filled band is $|d_0+p_0\rangle$ at the $\Gamma$ point, the states in the high-energy bands are consisting of $|d_{\pm2}\rangle$. 
The high-energy bands have to be degenerated at the $\Gamma$ point because of the mirror symmetry in the $y$ axis (see Fig.\ \ref{fig_schematic}).
The $\Gamma$ point is an invariant momentum of this mirror operation, which replaces the $d_2$ and $d_{-2}$ orbitals.
Therefore the $k$-independent term of Hamiltonian is given by 
\begin{align}
H_\Gamma(0)=E_0I_{3\times3}+\mathrm{diag}[0,E_b,E_b], 
\end{align}
where $E_0$ is the energy of the partially-filled band at the $\Gamma$ point, $I_{3\times3}$ is the identity matrix, and the basis is $(c_{d_0+p_0}, c_{d_{-2}},c_{d_{2}})$.

\begin{widetext}
The full effective Hamiltonian including the $k$-dependent terms is represented by 
\begin{align}
H_\Gamma(\boldsymbol{k})=(E_0-ak^2)I_{3\times3}
+\begin{pmatrix}
0&ivke^{-i\theta_k}-wk^2e^{2i\theta_k}&ivke^{i\theta_k}-wk^2e^{-2i\theta_k}\\
-ivke^{i\theta_k}-wk^2e^{-2i\theta_k}&E_b&0\\
-ivke^{-i\theta_k}-wk^2e^{2i\theta_k}&0&E_b
\end{pmatrix},\label{eq_hamiltonain_gamma}
\end{align}
with $\theta_k=\arctan(k_y/k_x)$.
\end{widetext}
The sign and the additional phase factor of $\pi/2$ in the $k$-linear terms are consistent with the phase difference in Fig.\ \ref{fig_orbitals}(c) and reflect three-fold rotation symmetry in the $z$ axis.
Under the rotation, the additional phase appearing in each off-diagonal element is canceled due to the extra phase appearing in the Wannier orbital with the non-zero angular momentum.
There is also a restriction due to time-reversal symmetry in the off-diagonal components, that is the same coefficient $v$ for the (1,2) and (1,3) components.
Since the $d_2$ and $d_{-2}$ orbitals have opposite angular momentum to each other, time-reversal operation exchanges these orbitals in the basis and its representation $\mathcal{T}$ is given as
\begin{align}
\mathcal{T}=i\mathcal{C}\begin{pmatrix}
1&0&0\\
0&0&1\\
0&1&0
\end{pmatrix},
\end{align}
with the complex conjugation operator $\mathcal{C}$.
To be time-reversal symmetric, the effective Hamiltonian fulfils $\mathcal{T}^\dagger H_\Gamma(\boldsymbol{k})\mathcal{T}=H_\Gamma(-\boldsymbol{k})$.
The Hamiltonian also preserves the mirror symmetry in the $y$ axis which exchanges these orbitals and changes the sign of $k_y$.

\begin{figure}[htbp]
\begin{center}
 \includegraphics[width=75mm]{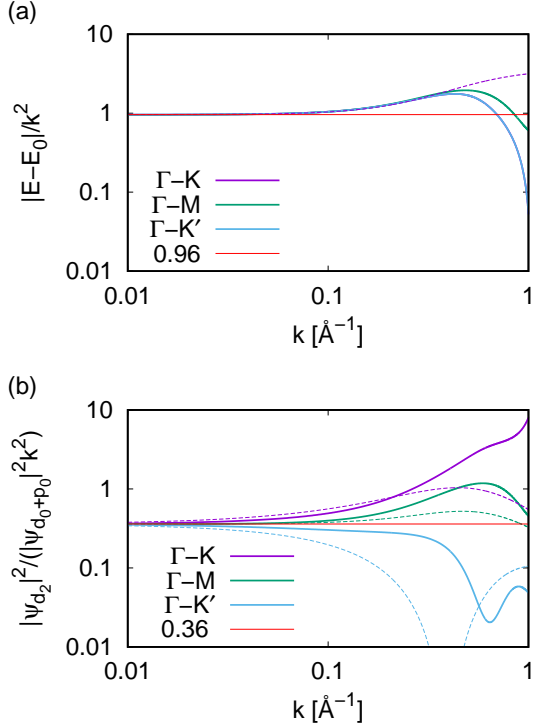}
\caption{
The wave vector dependence of energy dispersion and orbital amplitude around the $\Gamma$ point. The solid and dashed lines are numerical results by using the first-principles calculation and the effective model, respectively. The distance between two high-symmetry points is $k=1.205$ \AA$^{-1}$ and $1.044$ \AA$^{-1}$ for the $\Gamma-K$ and $\Gamma-M$ lines, respectively.
 }\label{fig_amplitude}
\end{center}
\end{figure} 
The effective Hamiltonian provides the handleable forms. 
The energy dispersion of the partially-filled band, the band crossing the Fermi energy, is represented by 
\begin{align}
E=E_0-ak^2+\frac{E_b}{2}-\sqrt{\left(\frac{E_b}{2}\right)^2+2(v^2k^2+w^2k^4)}.\label{eq_dispersion_G}
\end{align}
The electronic states are also given by
\begin{align}
\psi_\Gamma(\boldsymbol{k})=\frac{1}{C}\begin{pmatrix}
1+\sqrt{1+2(\tilde{v}^2k^2+\tilde{w}^2k^4)}\\
-ike^{i\theta_k}(\tilde{v}-i\tilde{w}ke^{-3i(\theta_k-\pi/6)})\\
-ike^{i\theta_k}(\tilde{v}-i\tilde{w}ke^{3i(\theta_k+\pi/6)})
\end{pmatrix},\label{eq_state_G}
\end{align}
with $\tilde{o}=o/E_b$ and the normalization factor $C$ independent of $\theta_k$.
Here the detailed calculation is presented in App.\ \ref{ap_gamma}.
The energy dispersion is isotropic, i.e., independent of $\theta_k$, but the orbital amplitude oscillates as a function of $\theta_k$ with preserving three-fold rotation symmetry.
Especially for the $d_{2}$ ($d_{-2}$) orbital, the amplitude oscillates as $c+\sin 3\theta_k$ ($c-\sin 3\theta_k$) with $c>1$ when the amplitude of $d_0+p_0$, the first component, is much larger than the other component.
Thus the effective Hamiltonian reproduces the oscillation of amplitude consistent with that in the case of the first-principles calculation in Fig.\ \ref{fig_orbitals}(a).

\begin{table}
\begin{center}
\begin{tabular}{c c c c c c}
\hline
\hline
&$E_0$&$E_b$&$a$&$v$&$w$\\ \hline
NbS$_2$\ &\ 0.93\ &\ 2.05 \ &\ -0.96 \ &\ 1.37 \ &\ 2.90 \ \\ 
NbSe$_2$\ &\ 0.38 \ &\ 1.89 \ &\ \ 2.32 \ &\ 1.13 \ &\ 2.90 \ \\ 
TaS$_2$\ &\ 1.06\ &\ 2.20 \ &\ -0.38\ &\ 1.63 \ &\ 3.50 \ \\  \hline \hline
\end{tabular}
\caption{
The parameter set for reproducing the energy dispersion and the electronic states around the $\Gamma$ point in metallic monolayer TMDCs. The parameters are defined in the proper unit: eV for $E_\alpha$, eV$\cdot$\AA\ for $v$, and eV$\cdot$\AA$^2$ for $a$ and $w$.
 }\label{table_parameters_G}
\end{center}
\end{table}
\begin{figure}[htbp]
\begin{center}
 \includegraphics[width=85mm]{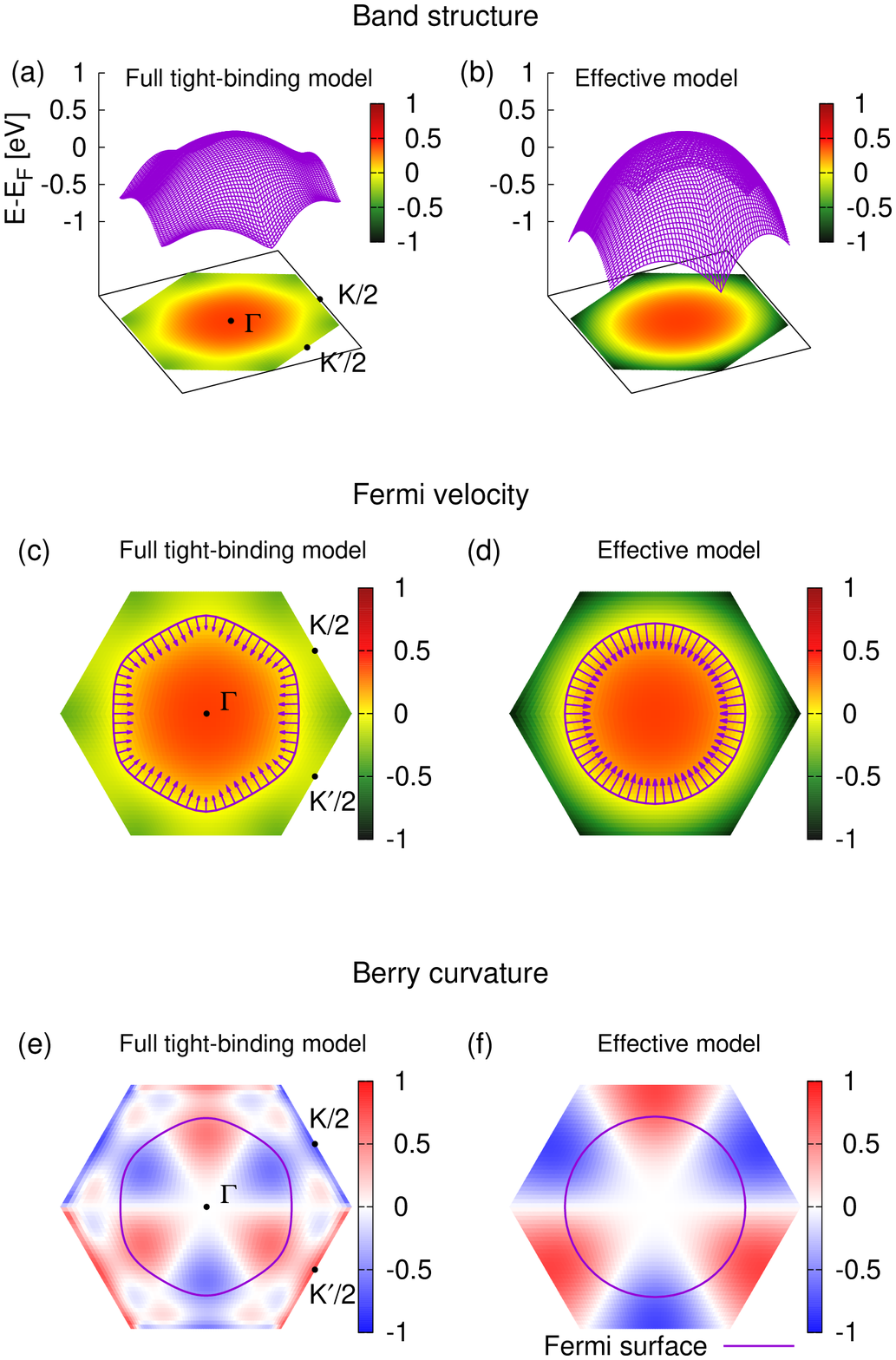}
\caption{
The three-dimensional band structure, the Fermi velocity, and the Berry curvature around the $\Gamma$ point. The left panels and right panels show the numerical results by using the multi-orbital tight-binding model and the three-orbital effective model, respectively. In (c) and (d), the Fermi velocity is indicated by arrows with the length in arbitrary units.  In (c), (d), (e), and (f), the Fermi pocket is depicted by a loop.  Here, $K/2$ and $K'/2$ indicate the middle points between the $\Gamma$-$K$ and the $\Gamma$-$K'$ lines, respectively. }\label{fig_3D_band_G}
\end{center}
\end{figure} 
We provide the parameters to reproduce the band structure and the electronic states in a realistic material for application in Table \ref{table_parameters_G}.
Monolayer NbSe$_2$ is an important material for application because several fascinating transport phenomena and the phase transitions have been observed experimentally in this material.\cite{Lu2015,Xi2015-super,Ugeda2015,Xi2015,Xi2017,Wang2017,Sohn2018,Zheng2018,Hamill2020}
Thus we demonstrate how to obtain these parameters from the first-principles results in the case of NbSe$_2$ for example.
At the  $\Gamma$ point, the constant coefficients can be estimated to be $E_0=0.32$ eV and $E_b=1.89$ eV as the maximum energy of the partially-filled band and the minimum energy of the $d_{\pm2}$-orbital-dominating bands, respectively.
For fitting the other parameters, we refer the $k$-dependence of energy dispersion and orbital-composition around the $\Gamma$ point as shown in Fig.\ \ref{fig_amplitude}(a) and (b), respectively.
In the limit of $k\rightarrow0$, the energy dispersion and the ratio of the $d_{\pm2}$ orbital amplitude to the $d_0+p_0$ orbital are represented by the asymptotic forms, $E_0-ak^2-2v^2k^2/E_b$ and $v^2k^2/E_b^2$, respectively. 
Then the coefficients of quadratic terms are estimated to be
\begin{align}
a+\frac{2v^2}{E_b} = 0.96\ \mathrm{eV}\cdot\mathrm{\AA}^2,\ \
\frac{v^2}{E_b^2}=0.36\ \mathrm{\AA}^2,
\end{align}
from the numerical results.
The another parameter $w$ is associated with the deviation from the quadratic dispersion and the trigonal oscillation of orbital amplitude.
We estimate this parameter $w=2.90$ eV$\cdot$\AA$^2$ by referring the deviation of energy dispersion from the quadratic form as shown in Fig.\ref{fig_amplitude}(a).

We test the validity of our model by calculating the three-dimensional band structure, the Fermi velocity, and the Berry curvature.
In Fig.\ \ref{fig_3D_band_G}, we show these quantities calculated by using the full tight-binding model from the first-principles band and the effective Hamiltonian.
The band structure and the Fermi velocity are well reproduced except the hexagonal warping of the Fermi surface.
More importantly, the effective model enable to reproduce the Berry curvature as shown in Fig.\ \ref{fig_3D_band_G}(f).
By using the conventional model, the electronic states in the $\Gamma$ valley is described by a single band Hamiltonian and their Berry curvature must be zero.
Therefore, our $3\times3$ effective model can be applied to the transport phenomena associated with the internal degree of freedom of electronic states, e.g., several Hall effects.

\subsection{The $K$ and $K'$ valleys }
In the $K$ ($K'$) valley, we adopt the $d_0$, $d_2$ ($d_{-2}$), $p_0$, and $p_1$ ($p_{-1}$) as the Wannier orbitals to constitute the basis.
We focus only on the effective model for the $K$ valley since the models in the two valleys can be replaced with each other by time-reversal operation $\mathcal{T}=i\mathcal{C}$ with $\boldsymbol{k}\rightarrow-\boldsymbol{k}$.
In terms of the $d_2$ and $p_1$ orbitals, the amplitude oscillates in the same way and the phase difference is a constant between them as shown in Fig.\ \ref{fig_orbitals} (b) and (d), respectively.
Thus we can introduce a composite orbital defined as
\begin{align}
|d_2+p_1\rangle=r_1'|d_2\rangle+r_2'|p_1\rangle, 
\end{align}
where the coefficients are given by $r_1'=\sqrt{0.83764}$ and $r_2'=-\sqrt{0.16236}\exp[-i\pi/6]$ in the case of NbSe$_2$.
Here the two orbitals $|d_2\rangle$ and $|p_1\rangle$ seem to obtain different phase factors under the three-fold rotation based only on their angular momenta but the relative atomic position causes these orbitals to obtain the same extra phase.
Therefore, the effective model can be represented by a $3\times3$ Hamiltonian defined on the basis of $(c_{d_0},\ c_{d_2+p_1},\ c_{p_0})$.

\begin{widetext}
The effective Hamiltonian for the $K$ valley is represented by 
\begin{align}
H_K(\boldsymbol{k})=&(E_0'-a'k^2)I_{3\times3}+r\begin{pmatrix}
E_b'&iu_1ke^{-i\theta_k}-w_1k^2e^{2i\theta_k}&0\\
-iu_1ke^{i\theta_k}-w_1k^2e^{-2i\theta_k}&0&iu_2ke^{i\theta_k}-w_2k^2e^{-2i\theta_k}\\
0&-iu_2ke^{-i\theta_k}-w_2k^2e^{2i\theta_k}&-E_b'
\end{pmatrix}.\label{eq_Hamiltonian_k}
\end{align}
\end{widetext}
Here we introduce two simplifications; the antisymmetry of upper and lower bands and the absence of coupling between the $d_0$ and $p_0$ orbitals because of the large gap between them.
These simplifications leads to no significant deterioration in the reproducibility of the partially-filled band and cause the Hamiltonian to be easily handleable.
The phase factor reflects the phase difference of realistic states in Fig.\ \ref{fig_orbitals}(d) and preserves three-fold rotation symmetry in the Hamiltonian.
\begin{figure}[htbp]
\begin{center}
 \includegraphics[width=80mm]{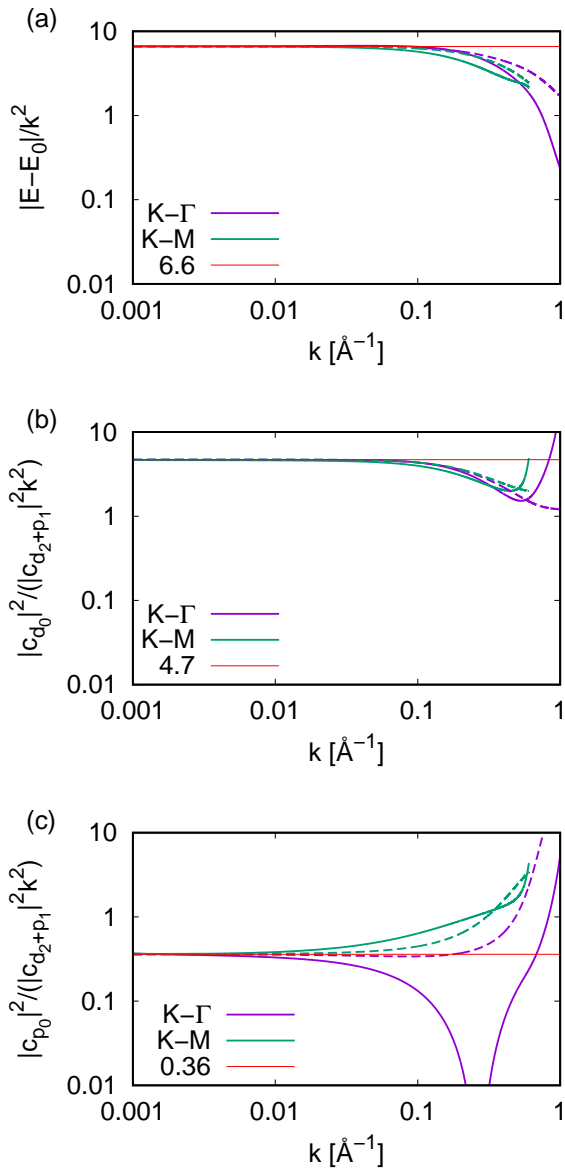}
\caption{
The variation of the energy dispersion and relative orbital amplitude. The solid and dashed lines represent the numerical results by using the first-principles calculation and the effective model, respectively. The distance between two high-symmetry points is $k=1.205$ \AA$^{-1}$ and $0.603$ \AA$^{-1}$ for the $K-\Gamma$ and $K-M$ lines, respectively.
 }\label{fig_amplitude_K}
\end{center}
\end{figure}

We present the analytic forms of energy dispersion and the electronic states by using the effective model.
Here the detailed calculation is provided in App.\ \ref{ap_k}.
Although the exact form of energy dispersion can be obtain (see Eq.\ (\ref{eq_exact_solution_k}) ), it is too complicated to analyze the qualitative property.
Thus we use the approximated form in Eq.\ (\ref{eq_approximated_form_k}) and present that of the partially-filled band,
\begin{align}
E=E_0'-a'k^2-E_b'\delta_-,
\end{align}
with
\begin{align}
\delta_\pm=\{&(u_1^2k^2+w_1^2k^4+2u_1w_1k^3\sin 3\theta_k)\nonumber\\&\pm(u_2^2k^2+w_2^2k^4-2u_2w_2k^3\sin 3\theta_k)\}/E_b'^2,\label{eq_delta}
\end{align}
under the condition that the gap energy $E_b'$ is much larger than the energy scales of other matrix components in the Hamiltonian.
The condition satisfied up to the Fermi wave number with respect to the $K$ point.
The electronic states are represented by the approximated form in Eq.\ (\ref{eq_approximated_state_k}),
\begin{align}
\psi_K(\boldsymbol{k})=\frac{1}{\sqrt{1+\delta_+^2/4}}\begin{pmatrix}
-ike^{i\theta_k}(\tilde{u}_1+i\tilde{w}_1k^2e^{-3i\theta_k})\\
1-\delta_+/2\\
-ike^{i\theta_k}(\tilde{u}_2-i\tilde{w}_2k^2e^{-3i\theta_k})
\end{pmatrix},
\end{align}
with $\tilde{o}=o/E_b'$.
Thus the $d_0$ and $p_0$ orbitals, the first and third components, oscillate as $1+c_1\sin 3\theta_k$ and $1-c_2\sin 3\theta_k$, respectively, with $c_j=u_jw_jk/(u_j^2+w_j^2k^2)$.
Moreover, the $d_2$ orbital is also oscillating as $1-c_0\sin 3\theta$ with $c_0=2(u_1w_1k^3-u_2w_2k^3)/(E_b'-u_1^2k^2-w_1^2k^4-u_2^2k^2-w_2^2k^4)$.
In NbSe$_2$, the oscillation of $|c_{d_0}|^2$ is much smaller than the others as shown in Fig.\ \ref{fig_amplitude} and thus one can set $w_1=0$.
This means that the trigonal warping effect can be included by only introducing the $p_0$ orbital in the effective model.

\begin{table}
\begin{center}
\begin{tabular}{c c c c c c c c c}
\hline
\hline
&$E_0'$&$E_b'$&$a'$&$u_1$&$u_2$&$w_1$&$w_2$&r\\ \hline
NbS$_2$&\ 0.56 \ &1.35 \ &0.16 \ &3.57 \ &0.91\ &0.0\ &1.3\ & 0.80\\ 
NbSe$_2$&\ 0.57 \ &1.45 \ &0.29 \ &3.15 \ &0.87\ &0.0\ &0.50\ & 0.78\\
TaS$_2$&\ 0.70 \ & 1.36 \ & 0.35 \ & 4.01 \ & 0.96 \ & 0.0 \ & 0.50 \ & 0.80 \\
 \hline \hline
\end{tabular}
\caption{
The parameter set for reproducing the energy dispersion and the electronic states around the $K$ point in monolayer NbSe$_2$. The parameters are defined in the proper unit: eV for $E'_\alpha$, eV$\cdot$\AA\ for $u_j$, and eV$\cdot$\AA$^2$ for $a'$ and $w_j$. Here, $r$ is a dimensionless parameter.
 }\label{table_parameters_K}
\end{center}
\end{table}
\begin{figure}[htbp]
\begin{center}
 \includegraphics[width=85mm]{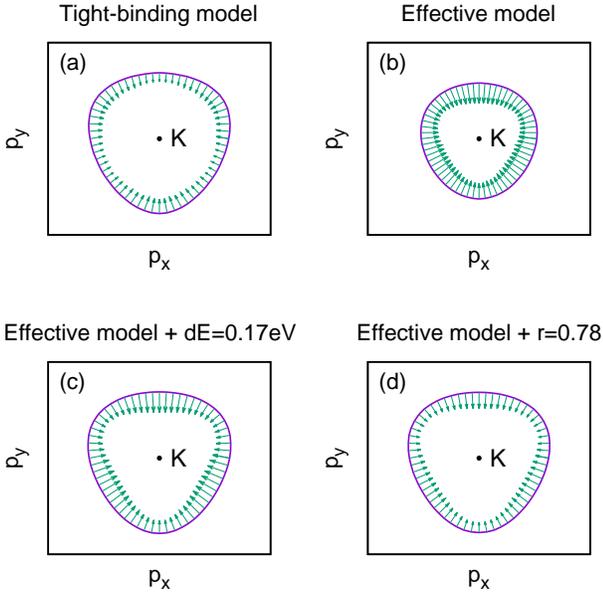}
\caption{
The Fermi velocity calculated in the different conditions. The loop represents the Fermi pocket around the $K$ point. In (a), both the Fermi velocity and pocket are obtained by using the tight-binding model based on the first-principles bands. The others are the numerical results by using the effective model.
}\label{fig_VF_K}
\end{center}
\end{figure} 
We provide the parameters for reproducing the electronic states in the $K$ valley of monolayer TMDCs in Table \ref{table_parameters_K} .
The estimation of the parameters is demonstrated in the case of NbSe$_2$ below.
The $k$-independent coefficients, $E_0'$ and $E_b'$, are determined from the first-principles energy eigenvalues of $E_0'=0.57eV$ and $E_b'=1.45$ eV, where the later refers to the bottom of the conduction band at the $K$ point.
The other parameters are estimated from the dispersion and the variation of orbital amplitude in electronic states.
We show the $k$-dependence of dispersion and orbital-amplitude around the $K$ point in Fig.\ \ref{fig_amplitude_K}.
The orbital-amplitude is given as the ratio to the $d_2+p_1$ orbital and it can be represented by the asymptotic forms $u_1^2k^2/E_b'^2$ for the $d_0$ orbital and $u_2^2k^2/E_b'^2$ for the $p_0$ orbital under $k\rightarrow0$.
The asymptotic form of dispersion is given by $-ak^2-(u_1^2-u_2^2)k^2/E_b'$ under $k\rightarrow0$.
By comparing with the first-principles calculation, we obtain
\begin{align}
\frac{u_1^2}{E_b'^2}=4.7\ \mathrm{\AA}^2,\ \ \frac{u_2^2}{E_b'^2}=0.36\ \mathrm{\AA}^2,\\ a+(u_1^2-u_2^2)/E_b'=6.6\ \mathrm{eV}\cdot\mathrm{\AA}^2.
\end{align}
The another parameters $w_j$ is associated with the trigonal warping effect (see Eq.\ (\ref{eq_delta})).
The trigonal warping term, which is proportional to $\sin3\theta_k$, leads to the difference of behavior along the two lines of $K-\Gamma$ and $K-M$.
In Fig.\ \ref{fig_amplitude_K}, the $k$-dependence is given in the lines in the first Brillouin zone.
For the $d_0$ orbital, such a difference is much smaller and indicates the negligibly small coefficient $w_1\sim0$.
On the other hand, for the $p_0$ orbital, there is a large trigonal warping effect which can be a reference to estimate $w_2$.
We adopt a small $w_2$ of 0.50 eV$\cdot$\AA$^2$ though it is insufficient for the reproducibility as shown in Fig.\ \ref{fig_SOC_K}(c).
This is because the appropriate parameter $w_2=3.3$ $\cdot$\AA$^2$ leads to the asymptotic behavior of dispersion $E(\boldsymbol{k})$ undesirable in the analysis, i.e., $E(\boldsymbol{k})\rightarrow\infty$ under $k\rightarrow\infty$ in the $K-M$ direction, and the mismatch in the Berry curvature.
We introduce another parameter $r$ for fitting the result of the effective model to that of the first-principles calculation as discussed below.

\begin{figure}[htbp]
\begin{center}
 \includegraphics[width=85mm]{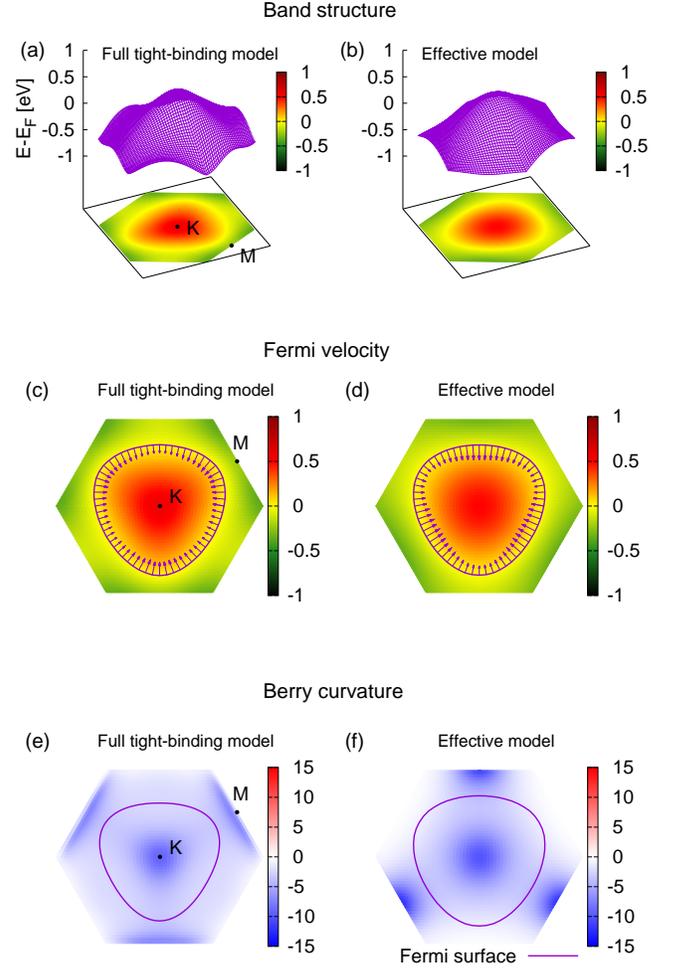}
\caption{
The three-dimensional band structure, the Fermi velocity, and the Berry curvature around the $K$ point. The left panels and right panels show the numerical results by using the multi-orbital tight-binding model and the three-orbital effective model, respectively. In (c) and (d), the Fermi velocity is indicated by arrows with the length in arbitrary units.  In (c), (d), (e), and (f), the Fermi pocket is depicted by a loop. }\label{fig_3D_band_K}
\end{center}
\end{figure} 
The parameter-set except $r$ is useful to reproduce the energy dispersion and electronic states near the $K$ point, but it is not sufficient to analyze those in the Fermi surface.
In Fig.\ \ref{fig_VF_K}, we show the Fermi surface and the Fermi velocity obtained by using different ways around the $K$ point.
In (a), these quantities are calculated by using the full tight-binding model from the first-principles band.
The result of the pristine effective model in (b) is much smaller than that of the tight-binding model. 
Thus we consider two modifications to the effective model: the shift of Fermi energy $dE$ in (c) and the scale change of band in (d).
Here the scale change is introduced by changing $r$ in Eq.\ \ref{eq_Hamiltonian_k} from unity.
The parameters for these modifications, $dE=0.17$eV and $r=0.78$, are adopted for the charge density, the area enclosed by the Fermi surface, to be equal to that in (a). 
In this paper, the scale change is adopted as the modification to the effective model because of the magnitude of Fermi velocity.

We confirm the validity of the effective model including the modification of $r$ by calculating the three-dimensional band structure and the Berry curvature in Fig.\ \ref{fig_3D_band_K}.
In the left panels and right panels, we present these quantities by using the the full tight-binding model from the first-principles band and the effective model, respectively.
The effective model can reproduce the band structure including the trigonal warping and the Berry curvature quantitatively without the increase around the $M$ point.

\section{Spin-orbit coupling}
In the electronic structure of TMDC, the SOC plays an important role due to inversion symmetry breaking. 
The SOC is represented by the conventional form\cite{Liu2013},
\begin{align}
H_{so}=\frac{\lambda}{\hbar^2}L_zs_z,
\end{align}
where $L_z$ and $s_z$ are the operators of the orbital and spin angular momenta perpendicular to the layer, respectively.
Here the parallel components are absent because of the crystal symmetry.
This SOC affects the band structure and the electronic states in the $\Gamma$ valley even though the conventional model cannot include the SOC effect in the $\Gamma$ valley.
The coupling constant $\lambda$ should be changed for fitting the electronic structure to the first-principles band and thus it takes different values in the $\Gamma$ and $K$ ($K'$) valleys.

\begin{figure}[htbp]
\begin{center}
 \includegraphics[width=85mm]{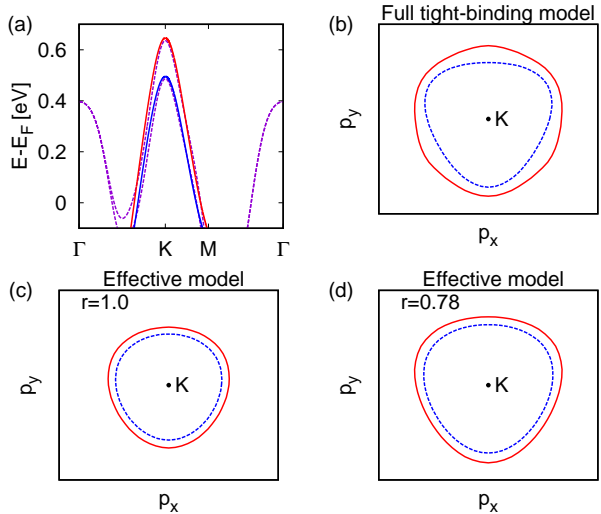}
\caption{The band structure and Fermi surface around the $K$ point in the presence of SOC. In (a), the first-principles band is depicted by the dashed line. In (b), (c), and (d), the solid line and the dashed line represents the up-spin and the down-spin bands, respectively.
 }\label{fig_SOC_K}
\end{center}
\end{figure} 
In the $K$ and $K'$ valleys, $\lambda$ can be estimated from the spin split $\Delta E_s$ at the $K$ and $K'$ points. 
The spin split is equal to the expectation value of the SOC operator,
\begin{align}
\Delta E_s=&2|\langle d_2+p_1|\frac{\lambda}{\hbar^2}L_zs_z|d_2+p_1\rangle|\nonumber\\
=&\lambda(2|r_1'|^2+|r_2'|^2)=0.150\ \mathrm{eV},
\end{align}
where the last quantity is obtained from the first-principles calculation in Fig.\ \ref{fig_band} (b).
Therefore the coupling constant is estimated to be
\begin{align}
\lambda=0.082\ \mathrm{eV}.
\end{align}
We show the band structure and the Fermi surface with SOC in Fig.\ \ref{fig_SOC_K}.
In the effective model, the SOC is represented by
\begin{align}
H_K^{\mathrm{so}}=\mathrm{diag}[0,\lambda(2|r_1'|^2+|r_2'|^2)/2,0],
\end{align}
and well describes the spin split band as shown in Fig.\ \ref{fig_SOC_K}(a).
The effective model reproduces the Fermi surface except for the hexagonal warping  in Fig\ \ref{fig_SOC_K}(b) and (d).
When one is interested in the conduction band, the $-\lambda |c_{p_{-1}}|^2/2$ with $|c_{p_{-1}}|^2=0.08$ is added as the (1,1) element of $H_K^{\mathrm{so}}$ as a contribution of mixed small amount of the $p_{-1}$ orbital in the $d_0$ orbital at the $K$ point.
It enable to reproduce the crossing of up-spin and down-spin branches in the conduction band.

\begin{figure}[htbp]
\begin{center}
 \includegraphics[width=85mm]{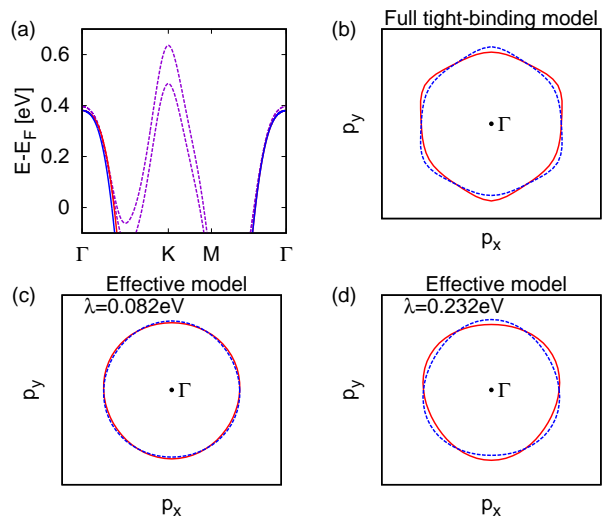}
\caption{The band structure and Fermi surface around the $\Gamma$ point in the presence of SOC. In (a), the first-principles band is depicted by the dashed line. In (b), (c), and (d), the solid line and the dashed line represents the up-spin and the down-spin bands, respectively.
 }\label{fig_SOC_G}
\end{center}
\end{figure} 
In the $\Gamma$ valley, SOC is represented by 
\begin{align}
H_\Gamma^{\mathrm{so}}=\mathrm{diag}[0,-\lambda',\lambda'].
\end{align}
We show the band structure and the Fermi surface in the presence of SOC in Fig.\ \ref{fig_SOC_G}.
The coupling constant should be larger than that estimated for the $K$ valley because the spin split is much smaller than that of the first-principles calculation in the case of $\lambda'=\lambda$ as shown in Fig.\ \ref{fig_SOC_G}(b) and (c).
We estimate the appropriate coupling constant of $\lambda'=0.23$ eV by fitting the maximum spin split between the up-spin and down-spin Fermi surfaces in Fig.\ \ref{fig_SOC_G}(b).
The effective model enables to reproduces the crossing between the up-spin and down-spin Fermi surface around the $\Gamma$ point.
The spin-dependent trigonal warping is attributed to the oscillation of $|c_{d_{\pm2}}|^2$ in Fig.\ \ref{fig_orbitals}(a).
In Table \ref{table_parameters_SOC}, we present the parameters to introduce SOC in some metallic TMDCs.
\begin{table}
\begin{center}
\begin{tabular}{c c c c c c c c c}
\hline
\hline
&$\lambda_\Gamma$&$\lambda_K$\\ \hline
NbS$_2$&\ 0.20 \ &0.065 \\
NbSe$_2$&\ 0.23 \ & 0.082 \\ 
TaS$_2$&\ 1.00 \ & 0.202  \\
 \hline \hline
\end{tabular}
\caption{
The coupling constant of SOC for the effective models at the $K$ and $K'$ valleys in TMDCs. The unit is eV.
 }\label{table_parameters_SOC}
\end{center}
\end{table}

\section{Conclusion}
We developed effective models in Eqs.\ (\ref{eq_hamiltonain_gamma}) and (\ref{eq_Hamiltonian_k}) for describing the electronic states in three valleys of metallic TMDCs. 
Every model is represented by $3\times3$ Hamiltonian which is continuous in the wave vector space and can be handled analytically.
The basis are consisting of not only the $d$ orbitals in transition-metal atoms but also the $p$ orbitals in chalcogen atoms.
Although the $p$ orbitals have been ignored in the conventional model for describing semiconducting TMDCs, we found that the $p_0$ orbital plays a crucial role for reproducing the energy dispersion and the electronic states in the metallic TMDCs.
We also reveals that the three-orbital model enables to analyze the internal degrees of freedom, e.g., the Berry curvature.
These models are applicable to electronic transport phenomena including complex orbital mixing effect, e.g., the transport in heterostructures with TMDCs.
Since these models also reproduce the phase structure of coefficient vector, this model can be applied to analyze the phase-related phenomena, e.g. several types of Hall effects.
We also provide the parameter set for reproducing the electronic states in NbSe$_2$, NbS$_2$, and TaS$_2$.

\begin{acknowledgements}
This work was supported by a Grant-in-Aid for Scientific Research on Innovative Areas "Topological Material Science" (KAKENHI Grant No. JP15H05852) from JSPS of Japan.
\end{acknowledgements}

\bibliography{TMDC}

\begin{thebibliography}{38}%
\makeatletter
\providecommand \@ifxundefined [1]{%
 \@ifx{#1\undefined}
}%
\providecommand \@ifnum [1]{%
 \ifnum #1\expandafter \@firstoftwo
 \else \expandafter \@secondoftwo
 \fi
}%
\providecommand \@ifx [1]{%
 \ifx #1\expandafter \@firstoftwo
 \else \expandafter \@secondoftwo
 \fi
}%
\providecommand \natexlab [1]{#1}%
\providecommand \enquote  [1]{``#1''}%
\providecommand \bibnamefont  [1]{#1}%
\providecommand \bibfnamefont [1]{#1}%
\providecommand \citenamefont [1]{#1}%
\providecommand \href@noop [0]{\@secondoftwo}%
\providecommand \href [0]{\begingroup \@sanitize@url \@href}%
\providecommand \@href[1]{\@@startlink{#1}\@@href}%
\providecommand \@@href[1]{\endgroup#1\@@endlink}%
\providecommand \@sanitize@url [0]{\catcode `\\12\catcode `\$12\catcode
  `\&12\catcode `\#12\catcode `\^12\catcode `\_12\catcode `\%12\relax}%
\providecommand \@@startlink[1]{}%
\providecommand \@@endlink[0]{}%
\providecommand \url  [0]{\begingroup\@sanitize@url \@url }%
\providecommand \@url [1]{\endgroup\@href {#1}{\urlprefix }}%
\providecommand \urlprefix  [0]{URL }%
\providecommand \Eprint [0]{\href }%
\providecommand \doibase [0]{http://dx.doi.org/}%
\providecommand \selectlanguage [0]{\@gobble}%
\providecommand \bibinfo  [0]{\@secondoftwo}%
\providecommand \bibfield  [0]{\@secondoftwo}%
\providecommand \translation [1]{[#1]}%
\providecommand \BibitemOpen [0]{}%
\providecommand \bibitemStop [0]{}%
\providecommand \bibitemNoStop [0]{.\EOS\space}%
\providecommand \EOS [0]{\spacefactor3000\relax}%
\providecommand \BibitemShut  [1]{\csname bibitem#1\endcsname}%
\let\auto@bib@innerbib\@empty
\bibitem [{\citenamefont {Naito}\ and\ \citenamefont
  {Tanaka}(1982)}]{Naito1982}%
  \BibitemOpen
  \bibfield  {author} {\bibinfo {author} {\bibfnamefont {M.}~\bibnamefont
  {Naito}}\ and\ \bibinfo {author} {\bibfnamefont {S.}~\bibnamefont {Tanaka}},\
  }\href {\doibase 10.1143/JPSJ.51.219} {\bibfield  {journal} {\bibinfo
  {journal} {Journal of the Physical Society of Japan}\ }\textbf {\bibinfo
  {volume} {51}},\ \bibinfo {pages} {219} (\bibinfo {year} {1982})}\BibitemShut
  {NoStop}%
\bibitem [{\citenamefont {Splendiani}\ \emph {et~al.}(2010)\citenamefont
  {Splendiani}, \citenamefont {Sun}, \citenamefont {Zhang}, \citenamefont {Li},
  \citenamefont {Kim}, \citenamefont {Chim}, \citenamefont {Galli},\ and\
  \citenamefont {Wang}}]{Splendiani2010}%
  \BibitemOpen
  \bibfield  {author} {\bibinfo {author} {\bibfnamefont {A.}~\bibnamefont
  {Splendiani}}, \bibinfo {author} {\bibfnamefont {L.}~\bibnamefont {Sun}},
  \bibinfo {author} {\bibfnamefont {Y.}~\bibnamefont {Zhang}}, \bibinfo
  {author} {\bibfnamefont {T.}~\bibnamefont {Li}}, \bibinfo {author}
  {\bibfnamefont {J.}~\bibnamefont {Kim}}, \bibinfo {author} {\bibfnamefont
  {C.-Y.}\ \bibnamefont {Chim}}, \bibinfo {author} {\bibfnamefont
  {G.}~\bibnamefont {Galli}}, \ and\ \bibinfo {author} {\bibfnamefont
  {F.}~\bibnamefont {Wang}},\ }\href {\doibase 10.1021/nl903868w} {\bibfield
  {journal} {\bibinfo  {journal} {Nano. Lett.}\ }\textbf {\bibinfo {volume}
  {10}},\ \bibinfo {pages} {1271} (\bibinfo {year} {2010})}\BibitemShut
  {NoStop}%
\bibitem [{\citenamefont {Lu}\ \emph {et~al.}(2015)\citenamefont {Lu},
  \citenamefont {Zheliuk}, \citenamefont {Leermakers}, \citenamefont {Yuan},
  \citenamefont {Zeitler}, \citenamefont {Law},\ and\ \citenamefont
  {Ye}}]{Lu2015}%
  \BibitemOpen
  \bibfield  {author} {\bibinfo {author} {\bibfnamefont {J.~M.}\ \bibnamefont
  {Lu}}, \bibinfo {author} {\bibfnamefont {O.}~\bibnamefont {Zheliuk}},
  \bibinfo {author} {\bibfnamefont {I.}~\bibnamefont {Leermakers}}, \bibinfo
  {author} {\bibfnamefont {N.~F.~Q.}\ \bibnamefont {Yuan}}, \bibinfo {author}
  {\bibfnamefont {U.}~\bibnamefont {Zeitler}}, \bibinfo {author} {\bibfnamefont
  {K.~T.}\ \bibnamefont {Law}}, \ and\ \bibinfo {author} {\bibfnamefont
  {J.~T.}\ \bibnamefont {Ye}},\ }\href {\doibase 10.1126/science.aab2277}
  {\bibfield  {journal} {\bibinfo  {journal} {Science}\ }\textbf {\bibinfo
  {volume} {350}},\ \bibinfo {pages} {1353} (\bibinfo {year}
  {2015})}\BibitemShut {NoStop}%
\bibitem [{\citenamefont {Xi}\ \emph {et~al.}(2015{\natexlab{a}})\citenamefont
  {Xi}, \citenamefont {Wang}, \citenamefont {Zhao}, \citenamefont {Park},
  \citenamefont {Law}, \citenamefont {Berger}, \citenamefont {Forro},
  \citenamefont {Shan},\ and\ \citenamefont {Mak}}]{Xi2015-super}%
  \BibitemOpen
  \bibfield  {author} {\bibinfo {author} {\bibfnamefont {X.}~\bibnamefont
  {Xi}}, \bibinfo {author} {\bibfnamefont {Z.}~\bibnamefont {Wang}}, \bibinfo
  {author} {\bibfnamefont {W.}~\bibnamefont {Zhao}}, \bibinfo {author}
  {\bibfnamefont {J.-H.}\ \bibnamefont {Park}}, \bibinfo {author}
  {\bibfnamefont {K.~T.}\ \bibnamefont {Law}}, \bibinfo {author} {\bibfnamefont
  {H.}~\bibnamefont {Berger}}, \bibinfo {author} {\bibfnamefont
  {L.}~\bibnamefont {Forro}}, \bibinfo {author} {\bibfnamefont
  {J.}~\bibnamefont {Shan}}, \ and\ \bibinfo {author} {\bibfnamefont {K.~F.}\
  \bibnamefont {Mak}},\ }\href {\doibase 10.1038/nphys3538} {\bibfield
  {journal} {\bibinfo  {journal} {Nature Physics}\ }\textbf {\bibinfo {volume}
  {12}},\ \bibinfo {pages} {139} (\bibinfo {year}
  {2015}{\natexlab{a}})}\BibitemShut {NoStop}%
\bibitem [{\citenamefont {Tang}\ \emph {et~al.}(2017)\citenamefont {Tang},
  \citenamefont {Zhang}, \citenamefont {Wong}, \citenamefont {Pedramrazi},
  \citenamefont {Tsai}, \citenamefont {Jia}, \citenamefont {Moritz},
  \citenamefont {Claassen}, \citenamefont {Ryu}, \citenamefont {Kahn} \emph
  {et~al.}}]{Tang2017}%
  \BibitemOpen
  \bibfield  {author} {\bibinfo {author} {\bibfnamefont {S.}~\bibnamefont
  {Tang}}, \bibinfo {author} {\bibfnamefont {C.}~\bibnamefont {Zhang}},
  \bibinfo {author} {\bibfnamefont {D.}~\bibnamefont {Wong}}, \bibinfo {author}
  {\bibfnamefont {Z.}~\bibnamefont {Pedramrazi}}, \bibinfo {author}
  {\bibfnamefont {H.-Z.}\ \bibnamefont {Tsai}}, \bibinfo {author}
  {\bibfnamefont {C.}~\bibnamefont {Jia}}, \bibinfo {author} {\bibfnamefont
  {B.}~\bibnamefont {Moritz}}, \bibinfo {author} {\bibfnamefont
  {M.}~\bibnamefont {Claassen}}, \bibinfo {author} {\bibfnamefont
  {H.}~\bibnamefont {Ryu}}, \bibinfo {author} {\bibfnamefont {S.}~\bibnamefont
  {Kahn}},  \emph {et~al.},\ }\href@noop {} {\bibfield  {journal} {\bibinfo
  {journal} {Nature Physics}\ }\textbf {\bibinfo {volume} {13}},\ \bibinfo
  {pages} {683} (\bibinfo {year} {2017})}\BibitemShut {NoStop}%
\bibitem [{\citenamefont {Xiao}\ \emph {et~al.}(2012)\citenamefont {Xiao},
  \citenamefont {Liu}, \citenamefont {Feng}, \citenamefont {Xu},\ and\
  \citenamefont {Yao}}]{Xiao2012}%
  \BibitemOpen
  \bibfield  {author} {\bibinfo {author} {\bibfnamefont {D.}~\bibnamefont
  {Xiao}}, \bibinfo {author} {\bibfnamefont {G.-B.}\ \bibnamefont {Liu}},
  \bibinfo {author} {\bibfnamefont {W.}~\bibnamefont {Feng}}, \bibinfo {author}
  {\bibfnamefont {X.}~\bibnamefont {Xu}}, \ and\ \bibinfo {author}
  {\bibfnamefont {W.}~\bibnamefont {Yao}},\ }\href {\doibase
  10.1103/PhysRevLett.108.196802} {\bibfield  {journal} {\bibinfo  {journal}
  {Phys. Rev. Lett.}\ }\textbf {\bibinfo {volume} {108}},\ \bibinfo {pages}
  {196802} (\bibinfo {year} {2012})}\BibitemShut {NoStop}%
\bibitem [{\citenamefont {Korm{\'a}nyos}\ \emph {et~al.}(2015)\citenamefont
  {Korm{\'a}nyos}, \citenamefont {Burkard}, \citenamefont {Gmitra},
  \citenamefont {Fabian}, \citenamefont {Z{\'o}lyomi}, \citenamefont
  {Drummond},\ and\ \citenamefont {Fal’ko}}]{Kormanyos2015}%
  \BibitemOpen
  \bibfield  {author} {\bibinfo {author} {\bibfnamefont {A.}~\bibnamefont
  {Korm{\'a}nyos}}, \bibinfo {author} {\bibfnamefont {G.}~\bibnamefont
  {Burkard}}, \bibinfo {author} {\bibfnamefont {M.}~\bibnamefont {Gmitra}},
  \bibinfo {author} {\bibfnamefont {J.}~\bibnamefont {Fabian}}, \bibinfo
  {author} {\bibfnamefont {V.}~\bibnamefont {Z{\'o}lyomi}}, \bibinfo {author}
  {\bibfnamefont {N.~D.}\ \bibnamefont {Drummond}}, \ and\ \bibinfo {author}
  {\bibfnamefont {V.}~\bibnamefont {Fal’ko}},\ }\href@noop {} {\bibfield
  {journal} {\bibinfo  {journal} {2D Materials}\ }\textbf {\bibinfo {volume}
  {2}},\ \bibinfo {pages} {022001} (\bibinfo {year} {2015})}\BibitemShut
  {NoStop}%
\bibitem [{\citenamefont {Cao}\ \emph {et~al.}(2012)\citenamefont {Cao},
  \citenamefont {Wang}, \citenamefont {Han}, \citenamefont {Ye}, \citenamefont
  {Zhu}, \citenamefont {Shi}, \citenamefont {Niu}, \citenamefont {Tan},
  \citenamefont {Wang}, \citenamefont {Liu},\ and\ \citenamefont
  {Feng}}]{Cao2012}%
  \BibitemOpen
  \bibfield  {author} {\bibinfo {author} {\bibfnamefont {T.}~\bibnamefont
  {Cao}}, \bibinfo {author} {\bibfnamefont {G.}~\bibnamefont {Wang}}, \bibinfo
  {author} {\bibfnamefont {W.}~\bibnamefont {Han}}, \bibinfo {author}
  {\bibfnamefont {H.}~\bibnamefont {Ye}}, \bibinfo {author} {\bibfnamefont
  {C.}~\bibnamefont {Zhu}}, \bibinfo {author} {\bibfnamefont {J.}~\bibnamefont
  {Shi}}, \bibinfo {author} {\bibfnamefont {Q.}~\bibnamefont {Niu}}, \bibinfo
  {author} {\bibfnamefont {P.}~\bibnamefont {Tan}}, \bibinfo {author}
  {\bibfnamefont {E.}~\bibnamefont {Wang}}, \bibinfo {author} {\bibfnamefont
  {B.}~\bibnamefont {Liu}}, \ and\ \bibinfo {author} {\bibfnamefont
  {J.}~\bibnamefont {Feng}},\ }\href {\doibase 10.1038/ncomms1882} {\bibfield
  {journal} {\bibinfo  {journal} {Nature Commun.}\ }\textbf {\bibinfo {volume}
  {3}},\ \bibinfo {pages} {887} (\bibinfo {year} {2012})}\BibitemShut {NoStop}%
\bibitem [{\citenamefont {Mak}\ \emph {et~al.}(2012)\citenamefont {Mak},
  \citenamefont {He}, \citenamefont {Shan},\ and\ \citenamefont
  {Heinz}}]{Mak2012}%
  \BibitemOpen
  \bibfield  {author} {\bibinfo {author} {\bibfnamefont {K.~F.}\ \bibnamefont
  {Mak}}, \bibinfo {author} {\bibfnamefont {K.}~\bibnamefont {He}}, \bibinfo
  {author} {\bibfnamefont {J.}~\bibnamefont {Shan}}, \ and\ \bibinfo {author}
  {\bibfnamefont {T.~F.}\ \bibnamefont {Heinz}},\ }\href {\doibase
  10.1038/nnano.2012.96} {\bibfield  {journal} {\bibinfo  {journal} {Nat Nano}\
  }\textbf {\bibinfo {volume} {7}},\ \bibinfo {pages} {494} (\bibinfo {year}
  {2012})}\BibitemShut {NoStop}%
\bibitem [{\citenamefont {Shi}\ \emph {et~al.}(2013)\citenamefont {Shi},
  \citenamefont {Pan}, \citenamefont {Zhang},\ and\ \citenamefont
  {Yakobson}}]{Shi2013}%
  \BibitemOpen
  \bibfield  {author} {\bibinfo {author} {\bibfnamefont {H.}~\bibnamefont
  {Shi}}, \bibinfo {author} {\bibfnamefont {H.}~\bibnamefont {Pan}}, \bibinfo
  {author} {\bibfnamefont {Y.-W.}\ \bibnamefont {Zhang}}, \ and\ \bibinfo
  {author} {\bibfnamefont {B.~I.}\ \bibnamefont {Yakobson}},\ }\href {\doibase
  10.1103/PhysRevB.87.155304} {\bibfield  {journal} {\bibinfo  {journal} {Phys.
  Rev. B}\ }\textbf {\bibinfo {volume} {87}},\ \bibinfo {pages} {155304}
  (\bibinfo {year} {2013})}\BibitemShut {NoStop}%
\bibitem [{\citenamefont {Habe}\ and\ \citenamefont
  {Koshino}(2017)}]{habe2017}%
  \BibitemOpen
  \bibfield  {author} {\bibinfo {author} {\bibfnamefont {T.}~\bibnamefont
  {Habe}}\ and\ \bibinfo {author} {\bibfnamefont {M.}~\bibnamefont {Koshino}},\
  }\href {\doibase 10.1103/PhysRevB.96.085411} {\bibfield  {journal} {\bibinfo
  {journal} {Phys. Rev. B}\ }\textbf {\bibinfo {volume} {96}},\ \bibinfo
  {pages} {085411} (\bibinfo {year} {2017})}\BibitemShut {NoStop}%
\bibitem [{\citenamefont {Korm\'anyos}\ \emph {et~al.}(2018)\citenamefont
  {Korm\'anyos}, \citenamefont {Z\'olyomi}, \citenamefont {Fal'ko},\ and\
  \citenamefont {Burkard}}]{Kormanyos2018}%
  \BibitemOpen
  \bibfield  {author} {\bibinfo {author} {\bibfnamefont {A.}~\bibnamefont
  {Korm\'anyos}}, \bibinfo {author} {\bibfnamefont {V.}~\bibnamefont
  {Z\'olyomi}}, \bibinfo {author} {\bibfnamefont {V.~I.}\ \bibnamefont
  {Fal'ko}}, \ and\ \bibinfo {author} {\bibfnamefont {G.}~\bibnamefont
  {Burkard}},\ }\href {\doibase 10.1103/PhysRevB.98.035408} {\bibfield
  {journal} {\bibinfo  {journal} {Phys. Rev. B}\ }\textbf {\bibinfo {volume}
  {98}},\ \bibinfo {pages} {035408} (\bibinfo {year} {2018})}\BibitemShut
  {NoStop}%
\bibitem [{\citenamefont {Shan}\ \emph {et~al.}(2013)\citenamefont {Shan},
  \citenamefont {Lu},\ and\ \citenamefont {Xiao}}]{Shan2013}%
  \BibitemOpen
  \bibfield  {author} {\bibinfo {author} {\bibfnamefont {W.-Y.}\ \bibnamefont
  {Shan}}, \bibinfo {author} {\bibfnamefont {H.-Z.}\ \bibnamefont {Lu}}, \ and\
  \bibinfo {author} {\bibfnamefont {D.}~\bibnamefont {Xiao}},\ }\href {\doibase
  10.1103/PhysRevB.88.125301} {\bibfield  {journal} {\bibinfo  {journal} {Phys.
  Rev. B}\ }\textbf {\bibinfo {volume} {88}},\ \bibinfo {pages} {125301}
  (\bibinfo {year} {2013})}\BibitemShut {NoStop}%
\bibitem [{\citenamefont {Ochoa}\ and\ \citenamefont
  {Rold\'an}(2013)}]{Ochoa2013-1}%
  \BibitemOpen
  \bibfield  {author} {\bibinfo {author} {\bibfnamefont {H.}~\bibnamefont
  {Ochoa}}\ and\ \bibinfo {author} {\bibfnamefont {R.}~\bibnamefont
  {Rold\'an}},\ }\href {\doibase 10.1103/PhysRevB.87.245421} {\bibfield
  {journal} {\bibinfo  {journal} {Phys. Rev. B}\ }\textbf {\bibinfo {volume}
  {87}},\ \bibinfo {pages} {245421} (\bibinfo {year} {2013})}\BibitemShut
  {NoStop}%
\bibitem [{\citenamefont {Ochoa}\ \emph {et~al.}(2013)\citenamefont {Ochoa},
  \citenamefont {Guinea},\ and\ \citenamefont {Fal'ko}}]{Ochoa2013-2}%
  \BibitemOpen
  \bibfield  {author} {\bibinfo {author} {\bibfnamefont {H.}~\bibnamefont
  {Ochoa}}, \bibinfo {author} {\bibfnamefont {F.}~\bibnamefont {Guinea}}, \
  and\ \bibinfo {author} {\bibfnamefont {V.~I.}\ \bibnamefont {Fal'ko}},\
  }\href {\doibase 10.1103/PhysRevB.88.195417} {\bibfield  {journal} {\bibinfo
  {journal} {Phys. Rev. B}\ }\textbf {\bibinfo {volume} {88}},\ \bibinfo
  {pages} {195417} (\bibinfo {year} {2013})}\BibitemShut {NoStop}%
\bibitem [{\citenamefont {Song}\ and\ \citenamefont {Dery}(2013)}]{Song2013}%
  \BibitemOpen
  \bibfield  {author} {\bibinfo {author} {\bibfnamefont {Y.}~\bibnamefont
  {Song}}\ and\ \bibinfo {author} {\bibfnamefont {H.}~\bibnamefont {Dery}},\
  }\href {\doibase 10.1103/PhysRevLett.111.026601} {\bibfield  {journal}
  {\bibinfo  {journal} {Phys. Rev. Lett.}\ }\textbf {\bibinfo {volume} {111}},\
  \bibinfo {pages} {026601} (\bibinfo {year} {2013})}\BibitemShut {NoStop}%
\bibitem [{\citenamefont {Hatami}\ \emph {et~al.}(2014)\citenamefont {Hatami},
  \citenamefont {Kernreiter},\ and\ \citenamefont {Z\"ulicke}}]{Hatami2014}%
  \BibitemOpen
  \bibfield  {author} {\bibinfo {author} {\bibfnamefont {H.}~\bibnamefont
  {Hatami}}, \bibinfo {author} {\bibfnamefont {T.}~\bibnamefont {Kernreiter}},
  \ and\ \bibinfo {author} {\bibfnamefont {U.}~\bibnamefont {Z\"ulicke}},\
  }\href {\doibase 10.1103/PhysRevB.90.045412} {\bibfield  {journal} {\bibinfo
  {journal} {Phys. Rev. B}\ }\textbf {\bibinfo {volume} {90}},\ \bibinfo
  {pages} {045412} (\bibinfo {year} {2014})}\BibitemShut {NoStop}%
\bibitem [{\citenamefont {Korm\'anyos}\ \emph {et~al.}(2014)\citenamefont
  {Korm\'anyos}, \citenamefont {Z\'olyomi}, \citenamefont {Drummond},\ and\
  \citenamefont {Burkard}}]{Kormanyos2014}%
  \BibitemOpen
  \bibfield  {author} {\bibinfo {author} {\bibfnamefont {A.}~\bibnamefont
  {Korm\'anyos}}, \bibinfo {author} {\bibfnamefont {V.}~\bibnamefont
  {Z\'olyomi}}, \bibinfo {author} {\bibfnamefont {N.~D.}\ \bibnamefont
  {Drummond}}, \ and\ \bibinfo {author} {\bibfnamefont {G.}~\bibnamefont
  {Burkard}},\ }\href {\doibase 10.1103/PhysRevX.4.011034} {\bibfield
  {journal} {\bibinfo  {journal} {Phys. Rev. X}\ }\textbf {\bibinfo {volume}
  {4}},\ \bibinfo {pages} {011034} (\bibinfo {year} {2014})}\BibitemShut
  {NoStop}%
\bibitem [{\citenamefont {Habe}\ and\ \citenamefont
  {Koshino}(2015)}]{Habe2015}%
  \BibitemOpen
  \bibfield  {author} {\bibinfo {author} {\bibfnamefont {T.}~\bibnamefont
  {Habe}}\ and\ \bibinfo {author} {\bibfnamefont {M.}~\bibnamefont {Koshino}},\
  }\href {\doibase 10.1103/PhysRevB.91.201407} {\bibfield  {journal} {\bibinfo
  {journal} {Phys. Rev. B}\ }\textbf {\bibinfo {volume} {91}},\ \bibinfo
  {pages} {201407(R)} (\bibinfo {year} {2015})}\BibitemShut {NoStop}%
\bibitem [{\citenamefont {Habe}\ and\ \citenamefont
  {Koshino}(2016)}]{Habe2016}%
  \BibitemOpen
  \bibfield  {author} {\bibinfo {author} {\bibfnamefont {T.}~\bibnamefont
  {Habe}}\ and\ \bibinfo {author} {\bibfnamefont {M.}~\bibnamefont {Koshino}},\
  }\href {\doibase 10.1103/PhysRevB.93.075415} {\bibfield  {journal} {\bibinfo
  {journal} {Phys. Rev. B}\ }\textbf {\bibinfo {volume} {93}},\ \bibinfo
  {pages} {075415} (\bibinfo {year} {2016})}\BibitemShut {NoStop}%
\bibitem [{\citenamefont {Xi}\ \emph {et~al.}(2016)\citenamefont {Xi},
  \citenamefont {Berger}, \citenamefont {Forr\'o}, \citenamefont {Shan},\ and\
  \citenamefont {Mak}}]{Xi2017}%
  \BibitemOpen
  \bibfield  {author} {\bibinfo {author} {\bibfnamefont {X.}~\bibnamefont
  {Xi}}, \bibinfo {author} {\bibfnamefont {H.}~\bibnamefont {Berger}}, \bibinfo
  {author} {\bibfnamefont {L.}~\bibnamefont {Forr\'o}}, \bibinfo {author}
  {\bibfnamefont {J.}~\bibnamefont {Shan}}, \ and\ \bibinfo {author}
  {\bibfnamefont {K.~F.}\ \bibnamefont {Mak}},\ }\href {\doibase
  10.1103/PhysRevLett.117.106801} {\bibfield  {journal} {\bibinfo  {journal}
  {Phys. Rev. Lett.}\ }\textbf {\bibinfo {volume} {117}},\ \bibinfo {pages}
  {106801} (\bibinfo {year} {2016})}\BibitemShut {NoStop}%
\bibitem [{\citenamefont {Wang}\ \emph {et~al.}(2017)\citenamefont {Wang},
  \citenamefont {Huang}, \citenamefont {Lin}, \citenamefont {Cui},
  \citenamefont {Chen}, \citenamefont {Zhu}, \citenamefont {Liu}, \citenamefont
  {Zeng}, \citenamefont {Zhou}, \citenamefont {Yu}, \citenamefont {Wang},
  \citenamefont {He}, \citenamefont {Tsang}, \citenamefont {Gao}, \citenamefont
  {Suenaga}, \citenamefont {Ma}, \citenamefont {Yang}, \citenamefont {Lu},
  \citenamefont {Yu}, \citenamefont {Teo}, \citenamefont {Liu},\ and\
  \citenamefont {Liu}}]{Wang2017}%
  \BibitemOpen
  \bibfield  {author} {\bibinfo {author} {\bibfnamefont {H.}~\bibnamefont
  {Wang}}, \bibinfo {author} {\bibfnamefont {X.}~\bibnamefont {Huang}},
  \bibinfo {author} {\bibfnamefont {J.}~\bibnamefont {Lin}}, \bibinfo {author}
  {\bibfnamefont {J.}~\bibnamefont {Cui}}, \bibinfo {author} {\bibfnamefont
  {Y.}~\bibnamefont {Chen}}, \bibinfo {author} {\bibfnamefont {C.}~\bibnamefont
  {Zhu}}, \bibinfo {author} {\bibfnamefont {F.}~\bibnamefont {Liu}}, \bibinfo
  {author} {\bibfnamefont {Q.}~\bibnamefont {Zeng}}, \bibinfo {author}
  {\bibfnamefont {J.}~\bibnamefont {Zhou}}, \bibinfo {author} {\bibfnamefont
  {P.}~\bibnamefont {Yu}}, \bibinfo {author} {\bibfnamefont {X.}~\bibnamefont
  {Wang}}, \bibinfo {author} {\bibfnamefont {H.}~\bibnamefont {He}}, \bibinfo
  {author} {\bibfnamefont {S.~H.}\ \bibnamefont {Tsang}}, \bibinfo {author}
  {\bibfnamefont {W.}~\bibnamefont {Gao}}, \bibinfo {author} {\bibfnamefont
  {K.}~\bibnamefont {Suenaga}}, \bibinfo {author} {\bibfnamefont
  {F.}~\bibnamefont {Ma}}, \bibinfo {author} {\bibfnamefont {C.}~\bibnamefont
  {Yang}}, \bibinfo {author} {\bibfnamefont {L.}~\bibnamefont {Lu}}, \bibinfo
  {author} {\bibfnamefont {T.}~\bibnamefont {Yu}}, \bibinfo {author}
  {\bibfnamefont {E.~H.~T.}\ \bibnamefont {Teo}}, \bibinfo {author}
  {\bibfnamefont {G.}~\bibnamefont {Liu}}, \ and\ \bibinfo {author}
  {\bibfnamefont {Z.}~\bibnamefont {Liu}},\ }\href {\doibase
  10.1038/s41467-017-00427-5} {\bibfield  {journal} {\bibinfo  {journal}
  {Nature Communications}\ }\textbf {\bibinfo {volume} {8}},\ \bibinfo {pages}
  {394} (\bibinfo {year} {2017})}\BibitemShut {NoStop}%
\bibitem [{\citenamefont {Sohn}\ \emph {et~al.}(2018)\citenamefont {Sohn},
  \citenamefont {Xi}, \citenamefont {He}, \citenamefont {Jiang}, \citenamefont
  {Wang}, \citenamefont {Kang}, \citenamefont {Park}, \citenamefont {Berger},
  \citenamefont {Forr{\'o}}, \citenamefont {Law} \emph {et~al.}}]{Sohn2018}%
  \BibitemOpen
  \bibfield  {author} {\bibinfo {author} {\bibfnamefont {E.}~\bibnamefont
  {Sohn}}, \bibinfo {author} {\bibfnamefont {X.}~\bibnamefont {Xi}}, \bibinfo
  {author} {\bibfnamefont {W.-Y.}\ \bibnamefont {He}}, \bibinfo {author}
  {\bibfnamefont {S.}~\bibnamefont {Jiang}}, \bibinfo {author} {\bibfnamefont
  {Z.}~\bibnamefont {Wang}}, \bibinfo {author} {\bibfnamefont {K.}~\bibnamefont
  {Kang}}, \bibinfo {author} {\bibfnamefont {J.-H.}\ \bibnamefont {Park}},
  \bibinfo {author} {\bibfnamefont {H.}~\bibnamefont {Berger}}, \bibinfo
  {author} {\bibfnamefont {L.}~\bibnamefont {Forr{\'o}}}, \bibinfo {author}
  {\bibfnamefont {K.~T.}\ \bibnamefont {Law}},  \emph {et~al.},\ }\href@noop {}
  {\bibfield  {journal} {\bibinfo  {journal} {Nat Mater}\ }\textbf {\bibinfo
  {volume} {17}},\ \bibinfo {pages} {504} (\bibinfo {year} {2018})}\BibitemShut
  {NoStop}%
\bibitem [{\citenamefont {Hamill}\ \emph {et~al.}(2020)\citenamefont {Hamill},
  \citenamefont {Heischmidt}, \citenamefont {Sohn}, \citenamefont {Shaffer},
  \citenamefont {Tsai}, \citenamefont {Zhang}, \citenamefont {Xi},
  \citenamefont {Suslov}, \citenamefont {Berger}, \citenamefont {Forr{\'o}}
  \emph {et~al.}}]{Hamill2020}%
  \BibitemOpen
  \bibfield  {author} {\bibinfo {author} {\bibfnamefont {A.}~\bibnamefont
  {Hamill}}, \bibinfo {author} {\bibfnamefont {B.}~\bibnamefont {Heischmidt}},
  \bibinfo {author} {\bibfnamefont {E.}~\bibnamefont {Sohn}}, \bibinfo {author}
  {\bibfnamefont {D.}~\bibnamefont {Shaffer}}, \bibinfo {author} {\bibfnamefont
  {K.-T.}\ \bibnamefont {Tsai}}, \bibinfo {author} {\bibfnamefont
  {X.}~\bibnamefont {Zhang}}, \bibinfo {author} {\bibfnamefont
  {X.}~\bibnamefont {Xi}}, \bibinfo {author} {\bibfnamefont {A.}~\bibnamefont
  {Suslov}}, \bibinfo {author} {\bibfnamefont {H.}~\bibnamefont {Berger}},
  \bibinfo {author} {\bibfnamefont {L.}~\bibnamefont {Forr{\'o}}},  \emph
  {et~al.},\ }\href@noop {} {\bibfield  {journal} {\bibinfo  {journal} {arXiv
  preprint arXiv:2004.02999}\ } (\bibinfo {year} {2020})}\BibitemShut {NoStop}%
\bibitem [{\citenamefont {Ugeda}\ \emph {et~al.}(2015)\citenamefont {Ugeda},
  \citenamefont {Bradley}, \citenamefont {Zhang}, \citenamefont {Onishi},
  \citenamefont {Chen}, \citenamefont {Ruan}, \citenamefont
  {Ojeda-Aristizabal}, \citenamefont {Ryu}, \citenamefont {Edmonds},
  \citenamefont {Tsai}, \citenamefont {Riss}, \citenamefont {Mo}, \citenamefont
  {Lee}, \citenamefont {Zettl}, \citenamefont {Hussain}, \citenamefont {Shen},\
  and\ \citenamefont {Crommie}}]{Ugeda2015}%
  \BibitemOpen
  \bibfield  {author} {\bibinfo {author} {\bibfnamefont {M.~M.}\ \bibnamefont
  {Ugeda}}, \bibinfo {author} {\bibfnamefont {A.~J.}\ \bibnamefont {Bradley}},
  \bibinfo {author} {\bibfnamefont {Y.}~\bibnamefont {Zhang}}, \bibinfo
  {author} {\bibfnamefont {S.}~\bibnamefont {Onishi}}, \bibinfo {author}
  {\bibfnamefont {Y.}~\bibnamefont {Chen}}, \bibinfo {author} {\bibfnamefont
  {W.}~\bibnamefont {Ruan}}, \bibinfo {author} {\bibfnamefont {C.}~\bibnamefont
  {Ojeda-Aristizabal}}, \bibinfo {author} {\bibfnamefont {H.}~\bibnamefont
  {Ryu}}, \bibinfo {author} {\bibfnamefont {M.~T.}\ \bibnamefont {Edmonds}},
  \bibinfo {author} {\bibfnamefont {H.-Z.}\ \bibnamefont {Tsai}}, \bibinfo
  {author} {\bibfnamefont {A.}~\bibnamefont {Riss}}, \bibinfo {author}
  {\bibfnamefont {S.-K.}\ \bibnamefont {Mo}}, \bibinfo {author} {\bibfnamefont
  {D.}~\bibnamefont {Lee}}, \bibinfo {author} {\bibfnamefont {A.}~\bibnamefont
  {Zettl}}, \bibinfo {author} {\bibfnamefont {Z.}~\bibnamefont {Hussain}},
  \bibinfo {author} {\bibfnamefont {Z.-X.}\ \bibnamefont {Shen}}, \ and\
  \bibinfo {author} {\bibfnamefont {M.~F.}\ \bibnamefont {Crommie}},\ }\href
  {\doibase 10.1038/nphys3527} {\bibfield  {journal} {\bibinfo  {journal}
  {Nature Physics}\ }\textbf {\bibinfo {volume} {12}},\ \bibinfo {pages} {92}
  (\bibinfo {year} {2015})}\BibitemShut {NoStop}%
\bibitem [{\citenamefont {Xi}\ \emph {et~al.}(2015{\natexlab{b}})\citenamefont
  {Xi}, \citenamefont {Zhao}, \citenamefont {Wang}, \citenamefont {Berger},
  \citenamefont {Forr\'o}, \citenamefont {Shan},\ and\ \citenamefont
  {Mak}}]{Xi2015}%
  \BibitemOpen
  \bibfield  {author} {\bibinfo {author} {\bibfnamefont {X.}~\bibnamefont
  {Xi}}, \bibinfo {author} {\bibfnamefont {L.}~\bibnamefont {Zhao}}, \bibinfo
  {author} {\bibfnamefont {Z.}~\bibnamefont {Wang}}, \bibinfo {author}
  {\bibfnamefont {H.}~\bibnamefont {Berger}}, \bibinfo {author} {\bibfnamefont
  {L.}~\bibnamefont {Forr\'o}}, \bibinfo {author} {\bibfnamefont
  {J.}~\bibnamefont {Shan}}, \ and\ \bibinfo {author} {\bibfnamefont {K.~F.}\
  \bibnamefont {Mak}},\ }\href {\doibase 10.1038/nnano.2015.143} {\bibfield
  {journal} {\bibinfo  {journal} {Nature Nanotechnology}\ }\textbf {\bibinfo
  {volume} {10}},\ \bibinfo {pages} {765} (\bibinfo {year}
  {2015}{\natexlab{b}})}\BibitemShut {NoStop}%
\bibitem [{\citenamefont {Zheng}\ \emph {et~al.}(2018)\citenamefont {Zheng},
  \citenamefont {Zhou}, \citenamefont {Liu},\ and\ \citenamefont
  {Feng}}]{Zheng2018}%
  \BibitemOpen
  \bibfield  {author} {\bibinfo {author} {\bibfnamefont {F.}~\bibnamefont
  {Zheng}}, \bibinfo {author} {\bibfnamefont {Z.}~\bibnamefont {Zhou}},
  \bibinfo {author} {\bibfnamefont {X.}~\bibnamefont {Liu}}, \ and\ \bibinfo
  {author} {\bibfnamefont {J.}~\bibnamefont {Feng}},\ }\href {\doibase
  10.1103/PhysRevB.97.081101} {\bibfield  {journal} {\bibinfo  {journal} {Phys.
  Rev. B}\ }\textbf {\bibinfo {volume} {97}},\ \bibinfo {pages} {081101}
  (\bibinfo {year} {2018})}\BibitemShut {NoStop}%
\bibitem [{\citenamefont {M\"ockli}\ and\ \citenamefont
  {Khodas}(2018)}]{David2018}%
  \BibitemOpen
  \bibfield  {author} {\bibinfo {author} {\bibfnamefont {D.}~\bibnamefont
  {M\"ockli}}\ and\ \bibinfo {author} {\bibfnamefont {M.}~\bibnamefont
  {Khodas}},\ }\href {\doibase 10.1103/PhysRevB.98.144518} {\bibfield
  {journal} {\bibinfo  {journal} {Phys. Rev. B}\ }\textbf {\bibinfo {volume}
  {98}},\ \bibinfo {pages} {144518} (\bibinfo {year} {2018})}\BibitemShut
  {NoStop}%
\bibitem [{\citenamefont {Aliabad}\ and\ \citenamefont
  {Zare}(2018)}]{Rahimi2018}%
  \BibitemOpen
  \bibfield  {author} {\bibinfo {author} {\bibfnamefont {M.~R.}\ \bibnamefont
  {Aliabad}}\ and\ \bibinfo {author} {\bibfnamefont {M.-H.}\ \bibnamefont
  {Zare}},\ }\href {\doibase 10.1103/PhysRevB.97.224503} {\bibfield  {journal}
  {\bibinfo  {journal} {Phys. Rev. B}\ }\textbf {\bibinfo {volume} {97}},\
  \bibinfo {pages} {224503} (\bibinfo {year} {2018})}\BibitemShut {NoStop}%
\bibitem [{\citenamefont {Habe}(2019{\natexlab{a}})}]{Habe2019-1}%
  \BibitemOpen
  \bibfield  {author} {\bibinfo {author} {\bibfnamefont {T.}~\bibnamefont
  {Habe}},\ }\href {\doibase 10.1063/1.5098802} {\bibfield  {journal} {\bibinfo
   {journal} {Journal of Applied Physics}\ }\textbf {\bibinfo {volume} {126}},\
  \bibinfo {pages} {123901} (\bibinfo {year} {2019}{\natexlab{a}})}\BibitemShut
  {NoStop}%
\bibitem [{\citenamefont {Habe}(2019{\natexlab{b}})}]{Habe2019-2}%
  \BibitemOpen
  \bibfield  {author} {\bibinfo {author} {\bibfnamefont {T.}~\bibnamefont
  {Habe}},\ }\href {\doibase 10.1103/PhysRevB.100.165431} {\bibfield  {journal}
  {\bibinfo  {journal} {Phys. Rev. B}\ }\textbf {\bibinfo {volume} {100}},\
  \bibinfo {pages} {165431} (\bibinfo {year} {2019}{\natexlab{b}})}\BibitemShut
  {NoStop}%
\bibitem [{\citenamefont {Glodzik}\ and\ \citenamefont
  {Ojanen}(2019)}]{Glodzik2019}%
  \BibitemOpen
  \bibfield  {author} {\bibinfo {author} {\bibfnamefont {S.}~\bibnamefont
  {Glodzik}}\ and\ \bibinfo {author} {\bibfnamefont {T.}~\bibnamefont
  {Ojanen}},\ }\href@noop {} {\bibfield  {journal} {\bibinfo  {journal}
  {arXiv:1905.01063}\ } (\bibinfo {year} {2019})}\BibitemShut {NoStop}%
\bibitem [{\citenamefont {Sticlet}\ and\ \citenamefont
  {Morari}(2019)}]{Sticlet2019}%
  \BibitemOpen
  \bibfield  {author} {\bibinfo {author} {\bibfnamefont {D.}~\bibnamefont
  {Sticlet}}\ and\ \bibinfo {author} {\bibfnamefont {C.}~\bibnamefont
  {Morari}},\ }\href {\doibase 10.1103/PhysRevB.100.075420} {\bibfield
  {journal} {\bibinfo  {journal} {Phys. Rev. B}\ }\textbf {\bibinfo {volume}
  {100}},\ \bibinfo {pages} {075420} (\bibinfo {year} {2019})}\BibitemShut
  {NoStop}%
\bibitem [{\citenamefont {Divilov}\ \emph {et~al.}(2020)\citenamefont
  {Divilov}, \citenamefont {Wan}, \citenamefont {Dreher}, \citenamefont
  {Ugeda},\ and\ \citenamefont {Yndur{\'a}in}}]{Divilov2020}%
  \BibitemOpen
  \bibfield  {author} {\bibinfo {author} {\bibfnamefont {S.}~\bibnamefont
  {Divilov}}, \bibinfo {author} {\bibfnamefont {W.}~\bibnamefont {Wan}},
  \bibinfo {author} {\bibfnamefont {P.}~\bibnamefont {Dreher}}, \bibinfo
  {author} {\bibfnamefont {M.~M.}\ \bibnamefont {Ugeda}}, \ and\ \bibinfo
  {author} {\bibfnamefont {F.}~\bibnamefont {Yndur{\'a}in}},\ }\href@noop {}
  {\bibfield  {journal} {\bibinfo  {journal} {arXiv preprint arXiv:2005.06210}\
  } (\bibinfo {year} {2020})}\BibitemShut {NoStop}%
\bibitem [{\citenamefont {Liu}\ \emph {et~al.}(2020)\citenamefont {Liu},
  \citenamefont {Chong}, \citenamefont {Sharma},\ and\ \citenamefont
  {Davis}}]{Liu2020}%
  \BibitemOpen
  \bibfield  {author} {\bibinfo {author} {\bibfnamefont {X.}~\bibnamefont
  {Liu}}, \bibinfo {author} {\bibfnamefont {Y.~X.}\ \bibnamefont {Chong}},
  \bibinfo {author} {\bibfnamefont {R.}~\bibnamefont {Sharma}}, \ and\ \bibinfo
  {author} {\bibfnamefont {J.}~\bibnamefont {Davis}},\ }\href@noop {}
  {\bibfield  {journal} {\bibinfo  {journal} {arXiv preprint arXiv:2007.15228}\
  } (\bibinfo {year} {2020})}\BibitemShut {NoStop}%
\bibitem [{\citenamefont {Giannozzi}\ \emph {et~al.}(2009)\citenamefont
  {Giannozzi}, \citenamefont {Baroni}, \citenamefont {Bonini}, \citenamefont
  {Calandra}, \citenamefont {Car}, \citenamefont {Cavazzoni}, \citenamefont
  {Ceresoli}, \citenamefont {Chiarotti}, \citenamefont {Cococcioni},
  \citenamefont {Dabo}, \citenamefont {Dal~Corso}, \citenamefont
  {de~Gironcoli}, \citenamefont {Fabris}, \citenamefont {Fratesi},
  \citenamefont {Gebauer}, \citenamefont {Gerstmann}, \citenamefont
  {Gougoussis}, \citenamefont {Kokalj}, \citenamefont {Lazzeri}, \citenamefont
  {Martin-Samos}, \citenamefont {Marzari}, \citenamefont {Mauri}, \citenamefont
  {Mazzarello}, \citenamefont {Paolini}, \citenamefont {Pasquarello},
  \citenamefont {Paulatto}, \citenamefont {Sbraccia}, \citenamefont {Scandolo},
  \citenamefont {Sclauzero}, \citenamefont {Seitsonen}, \citenamefont
  {Smogunov}, \citenamefont {Umari},\ and\ \citenamefont
  {Wentzcovitch}}]{quantum-espresso}%
  \BibitemOpen
  \bibfield  {author} {\bibinfo {author} {\bibfnamefont {P.}~\bibnamefont
  {Giannozzi}}, \bibinfo {author} {\bibfnamefont {S.}~\bibnamefont {Baroni}},
  \bibinfo {author} {\bibfnamefont {N.}~\bibnamefont {Bonini}}, \bibinfo
  {author} {\bibfnamefont {M.}~\bibnamefont {Calandra}}, \bibinfo {author}
  {\bibfnamefont {R.}~\bibnamefont {Car}}, \bibinfo {author} {\bibfnamefont
  {C.}~\bibnamefont {Cavazzoni}}, \bibinfo {author} {\bibfnamefont
  {D.}~\bibnamefont {Ceresoli}}, \bibinfo {author} {\bibfnamefont {G.~L.}\
  \bibnamefont {Chiarotti}}, \bibinfo {author} {\bibfnamefont {M.}~\bibnamefont
  {Cococcioni}}, \bibinfo {author} {\bibfnamefont {I.}~\bibnamefont {Dabo}},
  \bibinfo {author} {\bibfnamefont {A.}~\bibnamefont {Dal~Corso}}, \bibinfo
  {author} {\bibfnamefont {S.}~\bibnamefont {de~Gironcoli}}, \bibinfo {author}
  {\bibfnamefont {S.}~\bibnamefont {Fabris}}, \bibinfo {author} {\bibfnamefont
  {G.}~\bibnamefont {Fratesi}}, \bibinfo {author} {\bibfnamefont
  {R.}~\bibnamefont {Gebauer}}, \bibinfo {author} {\bibfnamefont
  {U.}~\bibnamefont {Gerstmann}}, \bibinfo {author} {\bibfnamefont
  {C.}~\bibnamefont {Gougoussis}}, \bibinfo {author} {\bibfnamefont
  {A.}~\bibnamefont {Kokalj}}, \bibinfo {author} {\bibfnamefont
  {M.}~\bibnamefont {Lazzeri}}, \bibinfo {author} {\bibfnamefont
  {L.}~\bibnamefont {Martin-Samos}}, \bibinfo {author} {\bibfnamefont
  {N.}~\bibnamefont {Marzari}}, \bibinfo {author} {\bibfnamefont
  {F.}~\bibnamefont {Mauri}}, \bibinfo {author} {\bibfnamefont
  {R.}~\bibnamefont {Mazzarello}}, \bibinfo {author} {\bibfnamefont
  {S.}~\bibnamefont {Paolini}}, \bibinfo {author} {\bibfnamefont
  {A.}~\bibnamefont {Pasquarello}}, \bibinfo {author} {\bibfnamefont
  {L.}~\bibnamefont {Paulatto}}, \bibinfo {author} {\bibfnamefont
  {C.}~\bibnamefont {Sbraccia}}, \bibinfo {author} {\bibfnamefont
  {S.}~\bibnamefont {Scandolo}}, \bibinfo {author} {\bibfnamefont
  {G.}~\bibnamefont {Sclauzero}}, \bibinfo {author} {\bibfnamefont {A.~P.}\
  \bibnamefont {Seitsonen}}, \bibinfo {author} {\bibfnamefont {A.}~\bibnamefont
  {Smogunov}}, \bibinfo {author} {\bibfnamefont {P.}~\bibnamefont {Umari}}, \
  and\ \bibinfo {author} {\bibfnamefont {R.~M.}\ \bibnamefont {Wentzcovitch}},\
  }\href {\doibase 10.1088/0953-8984/21/39/395502} {\bibfield  {journal}
  {\bibinfo  {journal} {J. Phys.: Condens. Matter}\ }\textbf {\bibinfo {volume}
  {21}},\ \bibinfo {pages} {395502} (\bibinfo {year} {2009})}\BibitemShut
  {NoStop}%
\bibitem [{\citenamefont {Mostofi}\ \emph {et~al.}(2008)\citenamefont
  {Mostofi}, \citenamefont {Yates}, \citenamefont {Lee}, \citenamefont {Souza},
  \citenamefont {Vanderbilt},\ and\ \citenamefont {Marzari}}]{wannier90}%
  \BibitemOpen
  \bibfield  {author} {\bibinfo {author} {\bibfnamefont {A.~A.}\ \bibnamefont
  {Mostofi}}, \bibinfo {author} {\bibfnamefont {J.~R.}\ \bibnamefont {Yates}},
  \bibinfo {author} {\bibfnamefont {Y.-S.}\ \bibnamefont {Lee}}, \bibinfo
  {author} {\bibfnamefont {I.}~\bibnamefont {Souza}}, \bibinfo {author}
  {\bibfnamefont {D.}~\bibnamefont {Vanderbilt}}, \ and\ \bibinfo {author}
  {\bibfnamefont {N.}~\bibnamefont {Marzari}},\ }\href {\doibase
  http://dx.doi.org/10.1016/j.cpc.2007.11.016} {\bibfield  {journal} {\bibinfo
  {journal} {Computer Physics Communications}\ }\textbf {\bibinfo {volume}
  {178}},\ \bibinfo {pages} {685 } (\bibinfo {year} {2008})}\BibitemShut
  {NoStop}%
\bibitem [{\citenamefont {Liu}\ \emph {et~al.}(2013)\citenamefont {Liu},
  \citenamefont {Shan}, \citenamefont {Yao}, \citenamefont {Yao},\ and\
  \citenamefont {Xiao}}]{Liu2013}%
  \BibitemOpen
  \bibfield  {author} {\bibinfo {author} {\bibfnamefont {G.-B.}\ \bibnamefont
  {Liu}}, \bibinfo {author} {\bibfnamefont {W.-Y.}\ \bibnamefont {Shan}},
  \bibinfo {author} {\bibfnamefont {Y.}~\bibnamefont {Yao}}, \bibinfo {author}
  {\bibfnamefont {W.}~\bibnamefont {Yao}}, \ and\ \bibinfo {author}
  {\bibfnamefont {D.}~\bibnamefont {Xiao}},\ }\href {\doibase
  10.1103/PhysRevB.88.085433} {\bibfield  {journal} {\bibinfo  {journal} {Phys.
  Rev. B}\ }\textbf {\bibinfo {volume} {88}},\ \bibinfo {pages} {085433}
  (\bibinfo {year} {2013})}\BibitemShut {NoStop}%
\end{thebibliography}%

\appendix
\begin{widetext}
\section{Effective model in the $\Gamma$ valley}\label{ap_gamma}
We introduced a $3\times3$ Hamiltonian to describe electronic states in the $\Gamma$ valley in the main text and represent it by
\begin{align}
H_\Gamma(\boldsymbol{k})
=&\begin{pmatrix}
0&ike^{-i\theta_k}(v+iwke^{3i\theta_k})&ike^{i\theta_k}(v+iw ke^{-3i\theta_k})\\
-ike^{i\theta_k}(v-iwke^{-3i\theta_k})&E_b&0\\
-ike^{-i\theta_k}(v-iwke^{3i\theta_k})&0&E_b
\end{pmatrix},
\end{align}
where the identity term $(E_0-ak^2)I_{3\times3}$ is omitted because it does not change the eigen vector.
Here we give the detailed calculation of the energy eigenvalue and the eigen vector in detail.
The energy eigenvalue can be calculated from the determinant,
\begin{align}
&\mathrm{det}[E-(H_\Gamma(\boldsymbol{k}))]\nonumber\\
=&E(E-E_b)^2-k^2|v+iwke^{3i\theta_k}|^2(E-E_b)-k^2|v+iwke^{-3i\theta_k}|^2(E-E_b)\nonumber\\
=&(E-E_b)\{E(E-E_b)-k^2(v^2+w^2k^2+iwvk(e^{3i\theta_3}-e^{-3i\theta_3}))-k^2(v^2+w^2k^2+iwvk(e^{-3i\theta_k}-e^{3i\theta_k})\}\nonumber\\
=&(E-E_b)\{E(E-E_b)-2k^2(v^2+w^2k^2)\}
.
\end{align}
The energy eigenvalue fulfills $\mathrm{det}[E-(H_\Gamma(\boldsymbol{k}))]=0$ and represented by
\begin{align}
E=E_b,\ \frac{E_b}{2}\pm\sqrt{\left(\frac{E_b}{2}\right)^2+2(v^2k^2+w^2k^4)}.
\end{align}
The wave function for the target band, $E=E_b/2-\sqrt{(E_b/2)^2+2(v^2k^2+w^2k^4)}$, is given by
\begin{align}
\psi_2(\boldsymbol{k})=\frac{1}{C}\begin{pmatrix}
1+\sqrt{1+2(\tilde{v}^2k^2+\tilde{w}^2k^4)}\\
-ike^{i\theta_k}(\tilde{v}-i\tilde{w}ke^{-3i(\theta_k-\pi/6)})\\
-ike^{i\theta_k}(\tilde{v}-i\tilde{w}ke^{3i(\theta_k+\pi/6)})
\end{pmatrix},
\end{align}
with $\tilde{o}=o/E_b'$ and the normalization factor,
\begin{align}
C^2=(1+\sqrt{1+2(\tilde{v}^2k^2+\tilde{w}^2k^4)})^2+2(\tilde{v}^2+\tilde{w}^2).
\end{align}

\section{Effective model in the K valley}\label{ap_k}
The introduced $3\times3$ Hamiltonian for the $K$ valley is given by
\begin{align}
H_K(\boldsymbol{k})=&\begin{pmatrix}
E_b'&iu_1ke^{i\theta_k}-w_1k^2e^{-2i\theta_k}&0\\
-iu_1ke^{-i\theta_k}-w_1k^2e^{2i\theta_k}&0&iu_2ke^{-i\theta_k}-w_2k^2e^{2i\theta_k}\\
0&-iu_2ke^{i\theta_k}-w_2k^2e^{-2i\theta_k}&-E_b'
\end{pmatrix}\\
=&\begin{pmatrix}
E_b'&ike^{i\theta_k}(u_1+iw_1ke^{-3i\theta_k})&0\\
-ike^{-i\theta_k}(u_1-iw_1ke^{3i\theta_k})&0&ike^{-i\theta}(u_2+iw_2ke^{3i\theta_k})\\
0&-ike^{i\theta}(u_2-iw_2ke^{-3i\theta_k})&-E_b'
\end{pmatrix}
.
\end{align}
The eigen energy can be calculated from the determinant as
\begin{align}
\mathrm{det}[E-H_K(\boldsymbol{k})]=&
(E-E_b')\{E(E+E_b')-k^2|u_2+iw_2ke^{3i\theta_k}|^2\}
-k^2|u_1+iw_1ke^{-3i\theta_k}|^2(E+E_b')\\
=&
E(E^2-E_b'^2)-k^2|u_2+iw_2ke^{3i\theta_k}|^2(E-E_b')
-k^2|u_1+iw_1ke^{-3i\theta_k}|^2(E+E_b')\\
=&
E(E^2-E_b'^2)-k^2(u_2^2+w_2^2k^2-2w_2u_2k\sin(3\theta_k))
(E-E_b')\nonumber\\
&-k^2(u_1^2+w_1^2k^2+2w_1u_1k\sin(3\theta_k))(E+E_b')\\
=&
E(E^2-E_b'^2)-k^2(u_1^2+u_2^2+(w_1^2+w_2^2)k^2+2(w_1u_1-w_2u_2)k\sin(3\theta_k))E\nonumber\\
&-E_bk^2\{(u_1^2+w_1^2k^2)-(u_2^2+w_2^2k^2)+2(w_1u_1k+w_2u_2k)\sin(3\theta_k)\}
.
\end{align}
This determinant equation is equivalent to a cubic equation of $E$ as 
\begin{align}
E^3+pE+q=0,
\end{align}
with 
\begin{align}
p=-E_b'^2-(u_1^2k^2+w_1^2k^4+2w_1u_1k^3\sin 3\theta_k )-(u_2^2k^2+w_2^2k^4-2w_2u_2k^3\sin 3\theta_k)\\
q=-E_b\{(u_1^2k^2+w_1^2k^4+2w_1u_1k^3\sin 3\theta_k)-(u_2^2k^2+w_2^2k^4-2w_2u_2k^3\sin 3\theta_k)\}.
\end{align}
We rewrite the parameters as
\begin{align}
p=-E_b'^2-E_1(k)^2-E_2(k)^2,\\
q=-E_b'(E_1(k)^2-E_2(k)^2),
\end{align}
with $E_1(k)^2=u_1^2k^2+w_1^2k^4+2u_1w_1k^3\sin 3\theta_k$ and $E_2(k)^2=u_2^2k^2+w_2^2k^4-2u_2w_2k^3\sin 3\theta_k$.
This kind of equation has a formal solution as
\begin{align}
E=\omega^l\sqrt[3]{-\frac{q}{2}+\sqrt{\left(\frac{q}{2}\right)^2+\left(\frac{p}{3}\right)^3}}
+\omega^{3-k}\sqrt[3]{-\frac{q}{2}-\sqrt{\left(\frac{q}{2}\right)^2+\left(\frac{p}{3}\right)^3}},
\end{align}
for $l=0$-$2$ with the cubic root $\omega^2=1$.
The explicit form of energy dispersion is given by
\begin{align}
E=&
\omega^l\sqrt[3]{\frac{E_b'(E_1^2-E_2^2)}{2}+\sqrt{\frac{E_b'^2(E_1^2-E_2^2)^2}{4}-\frac{(E_b'^2+E_1^2+E_2^2)^3}{27}}}\\
&+
\omega^{3-k}\sqrt[3]{\frac{E_b'(E_1^2-E_2^2)}{2}-\sqrt{\frac{E_b'^2(E_1^2-E_2^2)^2}{4}-\frac{(E_b'^2+E_1^2+E_2^2)^3}{27}}}\\
=&
\omega^lE_b'\sqrt[3]{\frac{(E_1^2-E_2^2)}{2E_b'^2}+\frac{i}{3\sqrt{3}}\sqrt{\left(1+\frac{(E_1^2+E_2^2)}{E_b'^2}\right)^3-\frac{27(E_1^2-E_2^2)^2}{4E_b'^4}}}\\
&+
\omega^{3-k}E_b'\sqrt[3]{\frac{(E_1^2-E_2^2)}{2E_b'^2}-\frac{i}{3\sqrt{3}}\sqrt{\left(1+\frac{E_1^2+E_2^2}{E_b'^2}\right)^3-\frac{27(E_1^2-E_2^2)^2}{4E_b'^4}}}\\
=&
-i\omega^l\frac{E_b'}{\sqrt{3}}\sqrt[3]{\frac{3\sqrt{3}(E_1^2-E_2^2)}{2iE_b'^2}+\sqrt{\left(1+\frac{(E_1^2+E_2^2)}{E_b'^2}\right)^3-\frac{27(E_1^2-E_2^2)^2}{4E_b'^4}}}\\
&+
i\omega^{3-k}\frac{E_b'}{\sqrt{3}}\sqrt[3]{-\frac{3\sqrt{3}(E_1^2-E_2^2)}{2iE_b'^2}+\sqrt{\left(1+\frac{E_1^2+E_2^2}{E_b'^2}\right)^3-\frac{27(E_1^2-E_2^2)^2}{4E_b'^4}}}\\
=&
-i\omega^l\frac{E_b'}{\sqrt{3}}\sqrt[3]{\frac{3\sqrt{3}}{2i}\delta_-+\sqrt{\left(1+\delta_+\right)^3-\frac{27}{4}\delta_-^2}}
+
i\omega^{3-k}\frac{E_b'}{\sqrt{3}}\sqrt[3]{-\frac{3\sqrt{3}}{2i}\delta_-+\sqrt{\left(1+\delta_+\right)^3-\frac{27}{4}\delta_-^2}}\label{eq_exact_solution_k}
,
\end{align}
where $\delta_\pm=(E_1^2\pm E_2^2)/E_b'^2$ is a much small value because of the large gap $E_b'$ up to the Fermi wave number.
At the $\Gamma$ point, $\delta_\pm$ goes to zero and the energy eigenvalue also goes to zero for $l=0$.
Therefore, the target band is represented by the form with $l=0$ and it can represented by $E=-E_b'\Delta(k)$.
The eigen states are represented by
\begin{align}
\psi_K=\frac{1}{C}
\begin{pmatrix}
-ike^{i\theta_k}(u_1+iw_1ke^{-3i\theta_k})(1-\Delta(k))\\
E_b'(1+\Delta(k))(1-\Delta(k))\\
-ike^{i\theta_k}(u_2-iw_2ke^{3i\theta_k})(1+\Delta(k)),
\end{pmatrix},
\end{align}
with the normalization factor,
\begin{align}
C^2=(u_1^2k^2+w_1^2k^4+2u_1w_1\sin3\theta_k)(1-\Delta(k))^2+(u_2^2k^2+w_2^2k^4+2u_2w_2\sin3\theta_k)(1+\Delta(k))^2+E_b'^2(1-\Delta(k)^2)^2.
\end{align}
This form is not easy to handle because of the complex form of $\Delta(k)$.

We write the eigenvalue and the eigenstate in the approximated forms as
\begin{align}
E\simeq&-i\omega^l\frac{E_b'}{\sqrt{3}}\sqrt[3]{\frac{3\sqrt{3}}{2i}\delta_-+1+\frac{3}{2}(\delta_++\delta_+^2)-\frac{27}{8}\delta_-^2}
+
i\omega^{3-k}\frac{E_b'}{\sqrt{3}}\sqrt[3]{-\frac{3\sqrt{3}}{2i}\delta_-+1+\frac{3}{2}(\delta_++\delta_+^2)-\frac{27}{8}\delta_-^2}\\
\simeq&
\omega^l\frac{E_b'}{\sqrt{3}}\left\{-\frac{\sqrt{3}}{2}\delta_--i\left(1+\frac{1}{2}(\delta_++\delta_+^2)-\frac{9}{8}\delta_-^2\right)\right\}
+
\omega^{3-k}\frac{E_b'}{\sqrt{3}}\left\{-\frac{\sqrt{3}}{2}\delta_-+i\left(1+\frac{1}{2}(\delta_++\delta_+^2)-\frac{9}{8}\delta_-^2\right)\right\}.
\end{align}
The three eigenvalues are obtained as 
\begin{align}
E=-E_b'\delta_-,\ \ E_b'\left\{\frac{\delta_-}{2}\pm\left(1+\frac{1}{2}(\delta_++\delta_+^2)-\frac{9}{8}\delta_-^2\right)\right\},\label{eq_approximated_form_k}
\end{align}
and the fist one is the target band crossing the Fermi level.
The wave function can be represented by
\begin{align}
\psi=\frac{1}{C'}\begin{pmatrix}
-ike^{i\theta_k}(u_1+iw_1k^2e^{-3i\theta_k})(1-\delta_-)\\
E_b'(1-\delta_-^2)\\
-ike^{i\theta_k}(u_2-iw_2k^2e^{-3i\theta_k})(1+\delta_-)
\end{pmatrix},
\end{align}
with the normalization factor,
\begin{align}
C'^2=&E_b'^2(1-2\delta_-^2+\delta_-^4)+(u_1^2k^2+w_1^2k^4+2u_1w_1\sin 3\theta_k)(1-2\delta_-+\delta_-^2)+(u_2^2k^2+w_2^2k^4-2u_2w_2\sin 3\theta_k)(1+2\delta_-+\delta_-^2)\\
=&
E_b'^2(1-2\delta_-^2+\delta_-^4)+E_b'^2\delta_+(1+\delta_-^2)-2\delta_-^2E_b'^2\\
=&
E_b'^2(1+\delta_+-4\delta_-^2+\delta_+\delta_-^2+\delta_-^4).
\end{align}
Thus we obtain the approximated eigen vector,
\begin{align}
\tilde{\psi}_K\simeq\frac{1}{\sqrt{1+\delta_+^2/4}}\begin{pmatrix}
-ike^{i\theta_k}(\tilde{u}_1+i\tilde{w}_1k^2e^{-3i\theta_k})\\
1-\delta_+/2\\
-ike^{i\theta_k}(\tilde{u}_2-i\tilde{w}_2k^2e^{-3i\theta_k})
\end{pmatrix},\label{eq_approximated_state_k}
\end{align}
up to $O(\delta_\pm)$ with $\tilde{o}=o/E_b'$.

\end{widetext}
\end{document}